%% file: main.tex
\documentclass[sigconf]{acmart}

\copyrightyear{2026}
\acmYear{2026}
\setcopyright{cc}
\setcctype{by}
\acmConference[ICMI '26]{INTERNATIONAL CONFERENCE ON MULTIMODAL INTERACTION}{October 05--09, 2026}{Napoli, Italy}
\acmBooktitle{INTERNATIONAL CONFERENCE ON MULTIMODAL INTERACTION (ICMI '26), October 05--09, 2026, Napoli, Italy}
\acmDOI{10.1145/3776574.3831135}
\acmISBN{979-8-4007-2318-6/2026/10}

\usepackage{booktabs}
\usepackage{graphicx}
\usepackage{amsmath}
\usepackage{multirow}
\usepackage{xcolor}
\usepackage{colortbl}
\definecolor{posgreen}{HTML}{d4edda}
\definecolor{negred}{HTML}{f8d7da}
\usepackage{subcaption}
\usepackage{enumitem}
\usepackage{placeins}
\usepackage{float}

\graphicspath{{figures/}}


\begin{document}

\title{Memory-Driven Self-Disclosure and Relational Turning Points:\\ A Longitudinal Multimodal Study of Human-AI Interaction}

\author{Ryuichi Sumida}
\authornote{Corresponding author.}
\email{sumida.ryuichi.65m@st.kyoto-u.ac.jp}
\affiliation{%
  \institution{Graduate School of Informatics, Kyoto University}
  \city{Kyoto}
  \country{Japan}
}
\affiliation{%
  \institution{Equmenopolis, Inc.}
  \city{Tokyo}
  \country{Japan}
}

\author{Mao Saeki}
\affiliation{%
  \institution{Waseda University}
  \city{Tokyo}
  \country{Japan}
}
\affiliation{%
  \institution{Equmenopolis, Inc.}
  \city{Tokyo}
  \country{Japan}
}

\author{Masaki Eguchi}
\affiliation{%
  \institution{Waseda University}
  \city{Tokyo}
  \country{Japan}
}
\affiliation{%
  \institution{Equmenopolis, Inc.}
  \city{Tokyo}
  \country{Japan}
}

\author{Sadahiro Yoshikawa}
\affiliation{%
  \institution{Equmenopolis, Inc.}
  \city{Tokyo}
  \country{Japan}
}

\author{Koji Inoue}
\affiliation{%
  \institution{Graduate School of Informatics, Kyoto University}
  \city{Kyoto}
  \country{Japan}
}

\author{Tatsuya Kawahara}
\affiliation{%
  \institution{School of Informatics, Kyoto University}
  \city{Kyoto}
  \country{Japan}
}

\author{Yoichi Matsuyama}
\affiliation{%
  \institution{Waseda University}
  \city{Tokyo}
  \country{Japan}
}
\affiliation{%
  \institution{Equmenopolis, Inc.}
  \city{Tokyo}
  \country{Japan}
}

\renewcommand{\shortauthors}{Sumida et al.}

\input{sections/00_abstract}

\maketitle

\input{sections/01_introduction}
\input{sections/02_related_work}

\input{sections/03_method}
\input{sections/04_results}
\input{sections/05_discussion}
\input{sections/06_conclusion}
\input{sections/07_safety_statement}
\input{sections/09_acknowledgment}

\bibliographystyle{ACM-Reference-Format}
\bibliography{references}

\newpage
\appendix
\input{sections/08_appendix}

\end{document}

%% file: sections/00_abstract.tex
\begin{abstract}
As conversational AI systems are designed for repeated use, a central question is how a series of interactions becomes a relationship.
We present a longitudinal multimodal study of a memory-augmented conversational agent (24 participants $\times$ 10 sessions), in which participants rated five relational constructs---familiarity, self-disclosure, perceived memory, conversational quality, and enjoyment---after each session.
Two complementary dynamics emerge.
First, conversational quality strongly shapes how enjoyable a session feels in the moment but does not carry forward across sessions, whereas perceived memory is relationally conditioned---predicted by prior relational state rather than reflecting system capability alone---and it shapes later enjoyment indirectly, via subsequent self-disclosure.
Second, relationships are punctuated by discrete turning points---crashes and surges---that are partially traceable in multimodal behavior and open different intervention windows: surges are more behaviorally detectable in the moment, enjoyment surges persist more reliably than enjoyment crashes recover, and some crashes are better forecast from person-specific behavioral drift than detected after they have already occurred.
Together, the findings suggest that longitudinal human-AI relationships are built through both slow accumulation and abrupt turning points.
\end{abstract}

\keywords{Longitudinal human-AI interaction; relational dynamics; relational shifts; crash-surge asymmetry; multimodal behavior analysis; conversational agents}

%% file: sections/01_introduction.tex
\section{Introduction}

Human relationships are inherently longitudinal: self-disclosure progressively broadens and deepens through reciprocal exchange~\cite{altman1973social}, and trust, rapport, and commitment develop through repeated interaction.
Yet, as surveys in social robotics~\cite{leite2013social} and conversational AI~\cite{brandtzaeg2022my} have noted, the vast majority of research on human-AI relational dynamics remains confined to single-session studies.
This leaves a basic question unresolved: when people return to the same agent day after day, what actually makes the interaction feel like an ongoing relationship rather than a sequence of isolated conversations?

We argue that this question has to be studied on two timescales.
At one timescale, relationships accumulate gradually: continuity across sessions may make users feel recognized, invite deeper self-disclosure, and sustain a sense of relational growth.
At another timescale, relationships are punctuated by abrupt turning points: a session where the interaction suddenly feels flat, alienating, or repetitive, or a session where the user unexpectedly feels genuine connection.

To study these two mechanisms together, we conducted a longitudinal multimodal study in which 24 participants interacted with \textit{InteLLA}, a memory-augmented voice agent, across 10 daily sessions.
After each session, participants rated five relational constructs: familiarity, self-disclosure comfort, perceived memory, conversational quality, and enjoyment.
We then analyzed both the temporal dependencies among these constructs and the behavioral traces of abrupt relational shifts using text, audio, and video features extracted from the interactions.

A practical challenge is that continuous rating prediction is dominated by temporal autocorrelation: exploratory analysis showed that the best predictor of session $t$ ratings is often session $t{-}1$ ratings.
This makes absolute trajectory prediction less informative about what changes the relationship.
We therefore complement construct-level temporal modeling with an event-based analysis of discrete relational shifts---crashes and surges---and ask whether these turning points are traceable in observable multimodal behavior.

The contributions of this work are twofold:
\begin{enumerate}[leftmargin=*, nosep]
    \item \textbf{Within-session quality is distinct from cross-session relational growth.} We disentangle what makes a session feel good from what appears to sustain the relationship across sessions. Within sessions, Conversational Quality dominates enjoyment, but it does not carry forward. Across sessions, Perceived Memory functions as a longitudinal bridge: it is associated with deeper next-session self-disclosure, and memory's link to later enjoyment is carried through self-disclosure. Perceived Memory is itself relationally conditioned---shaped by prior Familiarity, Enjoyment, and self-disclosure---suggesting that it reflects how users experience continuity, not merely a system capability.
    \item \textbf{Crashes and surges call for asymmetric intervention strategies.} We operationalize crashes and surges in longitudinal human-AI interaction and find that positive and negative shifts are not equally observable or actionable. Surges are more detectable than crashes from same-session multimodal behavior, whereas some crashes are better anticipated from prior-session behavioral drift than detected in the moment. This points to a concrete design implication: adaptive agents need both cross-session drift monitoring for crash prevention and in-session recognition for surge reinforcement.
\end{enumerate}

Our analysis code is publicly available.\footnote{\url{https://github.com/ryuichi-sumida/relational-turning-points}}

%% file: sections/02_related_work.tex
\section{Related Work}

\subsection{Relational Dynamics in Longitudinal Human-Agent Interaction}

Prior work has modeled session-level rapport, engagement, and enjoyment from multimodal behavioral cues \cite{tickle1990nature,zhao2016rapport,muller2018detecting,matsuyama2016socially,bohus2009models,sidner2005explorations,oertel2020engagement,santana2025speechtojoy}.
However, these efforts largely treat relational states as properties of a single session rather than repeated interaction.

Longitudinal human-agent interaction research has identified novelty decay and personalization as critical challenges \cite{leite2013social,kanda2007two}, and work in conversational AI has shown that perceptions of a chatbot as a ``friend'' develop over weeks of use \cite{brandtzaeg2022my}.
Yet these longitudinal studies have relied almost exclusively on descriptive analyses of self-report questionnaires or behavioral frequency counts, without modeling temporal dynamics between constructs.
Classical relationship theories suggest such modeling is needed: Social Penetration Theory \cite{altman1973social} posits that relationships deepen through progressive self-disclosure driven by a cost-reward calculus, while Uncertainty Reduction Theory \cite{berger1975uncertainty} predicts that information-seeking declines rapidly in early interactions before giving way to reciprocity and liking.
Together, these frameworks imply that different relational dimensions---surface-level familiarity, disclosure depth, cognitive appraisal---may follow qualitatively different temporal profiles rather than a single unified trajectory.

Emerging evidence supports this heterogeneity: longitudinal studies have documented non-monotonic relationship trajectories modulated by perceived rewards and costs~\cite{skjuve2022longitudinal}, positive feedback loops between disclosure and relational appraisal~\cite{laban2023longterm}, the role of agent memory in facilitating disclosure~\cite{jo2024longtermmemory}, and self-disclosure effects in human-chatbot conversations~\cite{ho2018psychological}.
However, these studies examine disclosure and memory in isolation; no prior work has characterized how multiple relational constructs co-evolve across repeated human-AI interactions---for instance, whether changes in one dimension (e.g., perceived memory) predict subsequent changes in another (e.g., self-disclosure).

We adopt these human--human relational frameworks as a \emph{testable scaffold} for structuring longitudinal human-AI dynamics, not as an assumption that people relate to agents exactly as they do to other people. Indeed, evidence suggests the two are not interchangeable: Hidalgo et al.~\cite{hidalgo2021judge} show that people judge machines and humans by systematically different standards, weighting intentions and outcomes differently across the two. Such asymmetries can shape how relational appraisals form with an agent, and we therefore treat human-AI-specific deviations as an empirical question---surfacing them in our data rather than presupposing equivalence.

\subsection{Relational Shifts and Crash-Surge Asymmetry}

Relationships are also shaped by discrete negative events and moments of recovery.
Prior work has examined ruptures in psychotherapy \cite{lipner2022rupture,tsakalidis2021patterns} and dialogue breakdowns \cite{higashinaka2016dialogue}, but not cross-session relational shifts in longitudinal human-AI interaction.

Relationship science provides strong theoretical grounding for an asymmetry between positive and negative relational events.
Baumeister et al.\ \cite{baumeister2001bad} showed that ``bad is stronger than good'' across domains of cognition, emotion, and social interaction: negative events are more salient, are processed more deeply, and exert a disproportionate influence on outcomes.
Gottman's cascade model \cite{gottman1994cascade} further demonstrated that relationship dissolution follows a predictable sequence of escalating negativity, suggesting that negative relational shifts may leave distinctive behavioral signatures---an expectation consistent with the broader negativity bias, but one that has not been tested against positive shifts in longitudinal human-AI interaction.

Existing work, however, focuses predominantly on detecting negative events---breakdowns, ruptures, trust violations---in isolation.
No work characterizes the asymmetry between positive and negative relational shifts in longitudinal human-AI interaction, nor investigates whether both leave distinct traces in multimodal behavioral signals.
The present study addresses this gap by jointly modeling \emph{crashes} (sharp negative shifts) and \emph{surges} (sharp positive shifts) across repeated sessions, examining whether they exhibit asymmetric predictability from visual, vocal and verbal cues.

%% file: sections/03_method.tex
\section{Method}
\label{sec:method}

\subsection{Study Design and Participants}
\label{ssec:study_design}

We conducted a longitudinal study in which $N{=}24$ university students interacted with a conversational AI agent across 10 daily sessions.
Participants were undergraduate or graduate students with English proficiency at CEFR C1 or above; all sessions were conducted in English.
Sessions were completed remotely on participants' own computers, with audio and video captured through their personal webcams; consequently, camera hardware and recording conditions varied across participants (a source of measurement noise we return to in Section~\ref{subsec:limitations}).

After each session, participants completed a 10-item post-session questionnaire on a 7-point Likert scale (1\,=\,strongly disagree, 7\,=\,strongly agree; full questionnaire in Appendix~\ref{app:questionnaire}), yielding a total of 2{,}270 individual ratings across the study (94.6\% completion; missing sessions were due to recording failures or connectivity issues on the participant's end).
The 10 items operationalize five relational constructs (Table~\ref{tab:constructs}), each computed as the mean of two items.
Constructs were selected to span distinct facets of human-AI relational development, from familiarity and self-disclosure willingness to perceived memory, conversational quality, and enjoyment.
Social Penetration, operationalized here as self-disclosure comfort, captures the core behavioral mechanism of Social Penetration Theory~\cite{altman1973social}; we use the terms ``social penetration'' and ``self-disclosure'' interchangeably throughout.
All constructs showed acceptable reliability and confirmatory factor analysis supported the five-factor structure (full psychometric validation in Appendix~\ref{app:psychometric}).

\begin{table}[t]
\centering
\small
\caption{Five relational constructs and questionnaire items. Each construct is the mean of two 7-point Likert items.}
\label{tab:constructs}
\begin{tabular}{@{} l p{4.8cm} @{}}
\toprule
\textbf{Construct} & \textbf{Items} \\
\midrule
Familiarity ($w_1$) & Q1: sense of familiarity \newline Q2: empathizes with feelings \\
Social Penetration ($w_2$) & Q3: comfortable with wide topics \newline Q4: comfortable with personal matters \\
Perceived Memory ($w_3$) & Q5: remembers past conversations \newline Q6: understands me \\
Conv.\ Quality ($w_4$) & Q7: feels natural \newline Q8: pleasant speaking \\
Enjoyment ($w_5$) & Q9: fun to talk \newline Q10: want to talk again \\
\bottomrule
\end{tabular}
\end{table}

Following the final session, 18 participants took part in semi-structured interviews capturing qualitative reflections on relationship development, memory experiences, and trust formation with the agent.
The study was approved by the Ethics Review Committee on Research with Human Subjects of Waseda University.

\subsection{The InteLLA System}
\label{ssec:InteLLA}

\begin{figure}[t]
\centering
\includegraphics[width=\columnwidth]{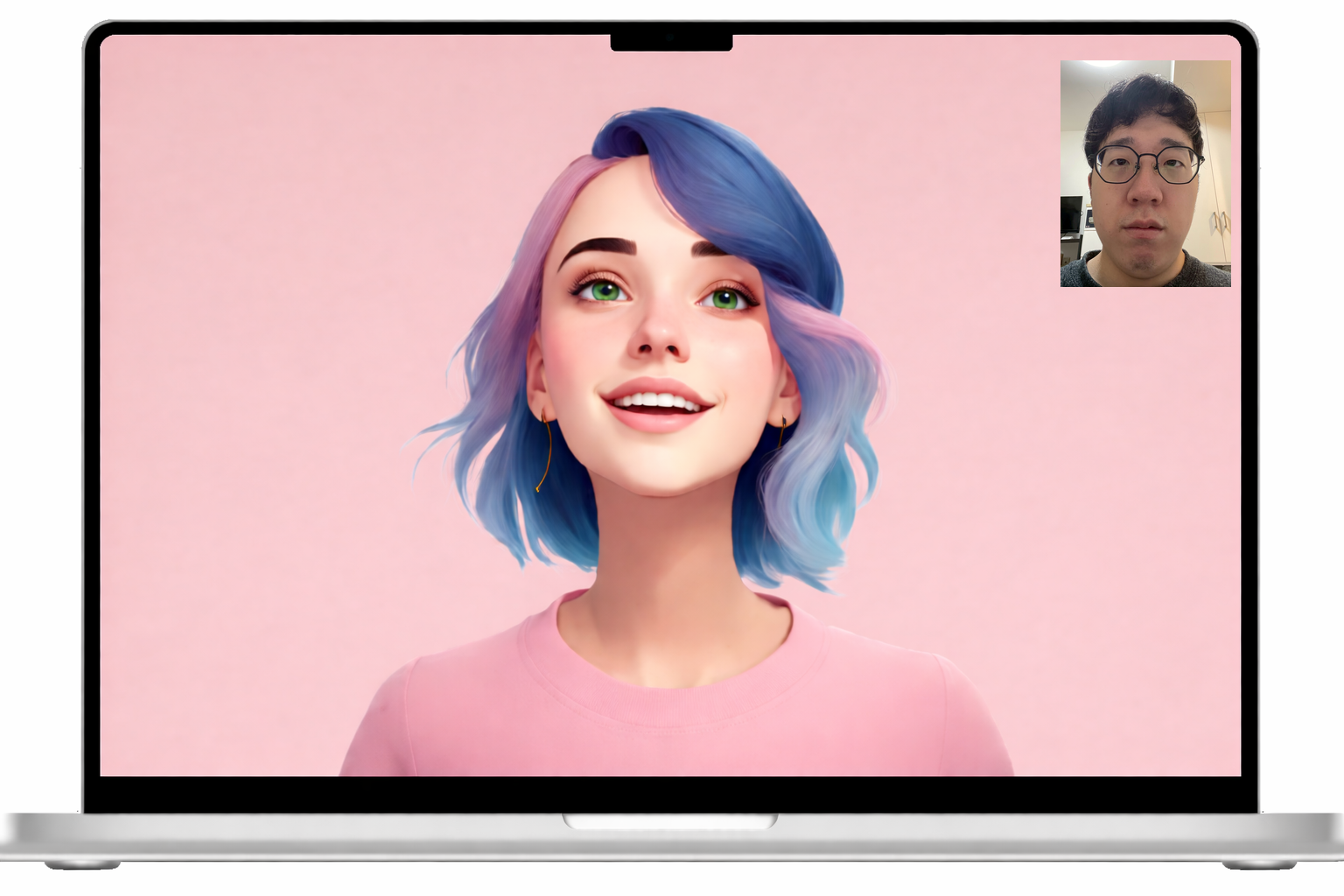}
\Description{Screenshot of the InteLLA conversational AI agent interface showing a text-based chat window with speech bubbles between the user and InteLLA, a cartoon avatar on the right side.}
\caption{The InteLLA conversational AI agent interface. Participants interacted with InteLLA through voice-based open-domain conversation across 10 daily sessions.}
\label{fig:InteLLA}
\end{figure}

InteLLA~\cite{saeki-etal-2024-InteLLA} is a voice-based conversational AI agent (Figure~\ref{fig:InteLLA}) powered by GPT-4o-mini, designed for friendly open-domain conversation.
Each session lasts approximately five minutes, comprising roughly 20--25 conversational turns.

To support cross-session continuity, InteLLA employs a retrieval-augmented generation (RAG) memory system~\cite{lewis2020retrieval} that retrieves relevant fragments from prior conversations at each turn,\allowbreak{} enabling the agent to reference past topics and personal details shared by the user.
After each session, a session summary is generated and appended to the user's memory store.
Before each session (except the first), a personalized ice-breaker is generated from the prior session's history, ensuring that each conversation opens with a reference to previously discussed content.

Sessions follow a three-phase structure: (1) a personalized opening referencing prior interactions, (2) open-domain chit-chat, and (3) a warm closing.
The full system prompt and memory-related prompts are provided in Appendix~\ref{app:agent_system_prompt}.

\subsection{Multimodal Feature Extraction}
\label{ssec:features}

We extract 351 interpretable features from three modalities---text, audio, and video---organized into 62 session-level features, 103 temporal features that capture longitudinal behavioral dynamics, and 186 person-normalized features that control for individual differences.
Table~\ref{tab:feature_summary} summarizes the feature inventory.

\begin{table}[t]
\centering
\small
\caption{Feature inventory by modality and temporal scope.}
\label{tab:feature_summary}
\begin{tabular}{@{} l r r r r @{}}
\toprule
\textbf{Modality} & \textbf{Session} & \textbf{Temporal} & \textbf{Person-Norm.} & \textbf{Total} \\
\midrule
Text   & 42 & 65 & 126 & 233 \\
Audio  & 16 & 26 &  48 &  90 \\
Video  &  4 & 12 &  12 &  28 \\
\midrule
\textbf{All} & \textbf{62} & \textbf{103} & \textbf{186} & \textbf{351} \\
\bottomrule
\end{tabular}
\end{table}

Text features (42 session-level) are derived via GPT-4.1~\cite{openai2025gpt41} turn-level annotation (emotion, vulnerability, disclosure markers, memory references, conversational acts) aggregated into nine conceptual families (full inventory in Table~\ref{tab:text_features}, Appendix~\ref{app:additional}); validation against two human raters shows moderate-to-substantial agreement (mean Cohen's $\kappa = 0.62$; Appendix~\ref{app:annotation_validation}).
Audio features (16 session-level) are extracted via openSMILE~\cite{eyben2010opensmile} (eGeMAPSv02~\cite{eyben2016geneva}), covering prosody, turn-taking, and emotion dynamics (Table~\ref{tab:audio_features}, Appendix~\ref{app:additional}).
Video features (4 session-level)---focused gaze percentage, attention switches, head motion energy, and blink rate---are extracted via OpenFace~2.0~\cite{baltrusaitis2018openface}; additional visual features (e.g., smile intensity, Action Unit activations) were excluded due to severe zero-inflation in voice-based interaction.
From these session-level features, we then derive 103 temporal features using four families of transformations---session-to-session deltas, max-prior values, exponential moving averages ($\alpha = 0.3$), and cumulative trend slopes---applied selectively based on each feature's measurement type and temporal properties.
We further derive 186 person-normalized features via participant deviation, $z$-score, and delta volatility transformations.
All prediction inputs are derived exclusively from observable interaction behaviors, with no access to self-report ratings.

\subsection{Temporal Dynamics Analysis}
\label{ssec:temporal_analysis}

We characterize the temporal structure of the five relational constructs using four complementary modeling approaches.
All models treat ratings as continuous and are estimated by ordinary least squares with participant fixed effects and participant-clustered standard errors.
In the notation below, $w_{k,i,t}$ is participant $i$'s rating of construct $k \in \{1,\dots,5\}$ at session $t \in \{1,\dots,10\}$, and $s_t = t - \bar{t}$ is the centered session index, with $\bar{t}$ the grand mean of all observed session indices ($\bar{t} \approx 5.3$).

\paragraph{Growth curve models}
For each construct $w_k$, we estimate a fixed-effects panel model
\begin{equation}
\label{eq:growth}
w_{k,i,t} = \alpha_{k,i} + \beta_k \, s_t + \epsilon_{k,i,t}
\end{equation}
where $\alpha_{k,i}$ are participant fixed effects and $\beta_k$ captures the average within-person linear trend across sessions.

\paragraph{Concurrent (same-session) associations}
To characterize the within-session relational structure, we regress each construct on the remaining four at the same time point:
\begin{equation}
\label{eq:concurrent}
w_{k,i,t} = \alpha_{k,i} + \sum_{j \neq k} \delta_{jk} \, w_{j,i,t} + \gamma_k \, s_t + \epsilon_{k,i,t}
\end{equation}
where $\delta_{jk}$ captures the partial association between constructs $j$ and $k$ within the same session, $\gamma_k$ is the coefficient on the centered session index $s_t$ (absorbing any linear temporal trend), $\alpha_{k,i}$ are participant fixed effects, and $\epsilon_{k,i,t}$ is the residual.
Cluster-robust standard errors are grouped by participant.
These same-session estimates serve as a baseline for comparison with the cross-lagged paths.

\paragraph{Fixed-effects cross-lagged panel model (CLPM)}
To test whether the level of one construct at session $t$ predicts another construct at session $t{+}1$, controlling for its autoregressive stability, we estimate pairwise cross-lagged regressions for all 20 directed pairs among the five constructs:
\begin{equation}
\label{eq:clpm}
w_{k,i,t+1} = \alpha_{k,i} + \beta_{j \to k} \, w_{j,i,t} + \rho_k \, w_{k,i,t} + \gamma_k \, s_t + \epsilon_{k,i,t}
\end{equation}
where $\beta_{j \to k}$ ($j \neq k$) is the cross-lagged coefficient of interest, $\rho_k$ controls for autoregressive stability, $\gamma_k$ is the coefficient on the centered session index $s_t$ (absorbing any linear temporal trend), $\alpha_{k,i}$ are participant fixed effects, and $\epsilon_{k,i,t}$ is the residual.
All models use cluster-robust standard errors grouped by participant.
Because the participant fixed effects $\alpha_{k,i}$ absorb time-invariant between-person differences, this specification is best interpreted as a within-person-oriented short-panel lagged regression rather than a pooled CLPM. As a further robustness check, we re-estimated the lagged models using person-mean centered variables (Appendix~\ref{app:robustness}); a full Random-Intercept CLPM would decompose stable trait variance from within-person dynamics more formally but requires a larger sample than our $N{=}24$.

\paragraph{Lagged mediation model}
To test whether Memory at session $t$ predicts Enjoyment at session $t{+}1$ indirectly through self-disclosure, we specify a mediation model with a lagged $a$-path and a contemporaneous $b$-path: Memory at session $t$ as predictor ($X$), Social Penetration at session $t{+}1$ as mediator ($M$), and Enjoyment at session $t{+}1$ as outcome ($Y$):
\begin{align}
\label{eq:med_a}
M_{i,t+1} &= \alpha^M_i + a \, X_{i,t} + \rho_M \, M_{i,t} + \gamma_M \, s_t + \epsilon^M_{i,t} \\
\label{eq:med_b}
Y_{i,t+1} &= \alpha^Y_i + b \, M_{i,t+1} + c' X_{i,t} + \rho_Y \, Y_{i,t} + \gamma_Y \, s_t + \epsilon^Y_{i,t}
\end{align}
The indirect effect $a \times b$ is tested via cluster bootstrap (5{,}000 iterations resampling at the participant level), with both percentile and bias-corrected confidence intervals.
As a robustness check, we also fit a parallel mediation model that includes both Social Penetration and Conversational Quality as simultaneous mediators, to confirm which channel carries the indirect effect.

\subsection{Crash and Surge Event Definitions}
\label{ssec:crash_surge}

We define two types of extreme relational events---\emph{crashes} and \emph{surges}---that capture abrupt deteriorations and improvements in the human-AI relationship between consecutive sessions.
These event definitions follow \citet{lipner2022rupture}, who operationalized clinically meaningful ruptures in therapeutic alliance using standard-deviation thresholds on session-to-session changes; we adopt the same 1-SD threshold, adapted here to the human-AI relational context (sensitivity analyses at 0.75 and 1.25 SD confirm that key patterns are preserved; Appendix~\ref{app:sensitivity}).

\paragraph{Event operationalization}
For each construct $w_k$ ($k \in \{1,\ldots,5\}$) and each session $t \in \{2,\ldots,10\}$, we compute the session-to-session delta
$\Delta w_k^{(t)} = w_k^{(t)} - w_k^{(t-1)}$
for each participant.
A \emph{crash} event at session $t$ is defined as $\Delta w_k^{(t)}$ more negative than one standard deviation below the mean delta for that construct, and a \emph{surge} is defined symmetrically as $\Delta w_k^{(t)}$ exceeding one standard deviation above the mean.

\paragraph{Leave-one-participant-out threshold computation}
To prevent data leakage, the mean and standard deviation used for threshold computation are calculated in a leave-one-participant-out (LOPO) fashion: when evaluating participant~$i$, the thresholds $\mu_{\Delta w_k}$ and $\sigma_{\Delta w_k}$ are computed from the remaining 23 participants' transitions only.
This ensures that no information from the held-out participant's own rating trajectory influences the event labels.
The LOPO thresholds exhibit high stability across folds, with coefficients of variation (CV) below 2.5\% for all constructs, confirming that no single participant disproportionately influences the threshold estimates.

\paragraph{Event rates and distributional properties}
Table~\ref{tab:event_counts} reports per-construct event counts and rates across the 202 session-to-session transitions.
Crash rates range from 8.9\% (Social Penetration, 18 events) to 17.3\% (Familiarity, 35 events); surge rates from 13.9\% (Enjoyment, 28) to 22.8\% (Conv.\ Quality, 46).
We additionally define \emph{systemic} events: a systemic crash (surge) occurs when two or more constructs crash (surge) simultaneously at the same transition, indicating a broad relational disruption (or breakthrough) rather than a construct-specific fluctuation.

\begin{table}[t]
\centering
\small
\caption{Event counts and distributional properties across 202 session-to-session transitions. Skew and kurtosis describe the delta ($\Delta w_k^{(t)}$) distribution per construct.}
\label{tab:event_counts}
\begin{tabular}{@{} l rr rr rr @{}}
\toprule
 & \multicolumn{2}{c}{\textbf{Crashes}} & \multicolumn{2}{c}{\textbf{Surges}} & & \\
\cmidrule(lr){2-3} \cmidrule(lr){4-5}
\textbf{Construct} & $n$ & \% & $n$ & \% & Skew & Kurt. \\
\midrule
Familiarity      & 35 & 17.3 & 34 & 16.8 & $-$0.01 & 0.51 \\
Social Penetr.   & 18 &  8.9 & 32 & 15.8 & $-$0.06 & 1.71 \\
Memory           & 24 & 11.9 & 29 & 14.4 &    0.27 & 1.00 \\
Conv.\ Quality   & 34 & 16.8 & 46 & 22.8 & $-$0.65 & 1.81 \\
Enjoyment        & 21 & 10.4 & 28 & 13.9 & $-$0.89 & 3.22 \\
\bottomrule
\end{tabular}
\end{table}

\begin{figure*}[t]
\centering
\begin{subfigure}[t]{0.49\textwidth}
\centering
\includegraphics[width=\textwidth]{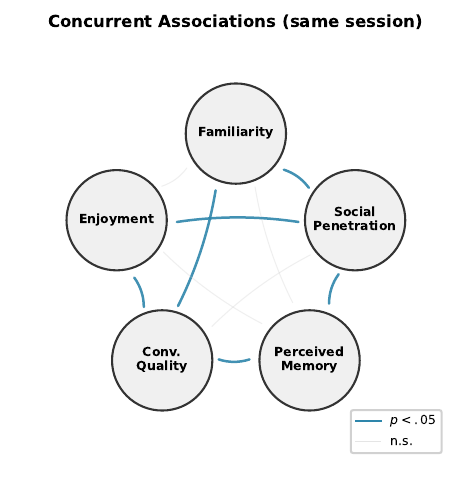}
\Description{Network diagram showing concurrent same-session associations among five relational constructs, with Conversational Quality showing the strongest edges to Enjoyment and Familiarity.}
\caption{Concurrent (same-session) associations. Conv.\ Quality dominates within sessions.}
\label{fig:concurrent_paths}
\end{subfigure}
\hfill
\begin{subfigure}[t]{0.49\textwidth}
\centering
\includegraphics[width=\textwidth]{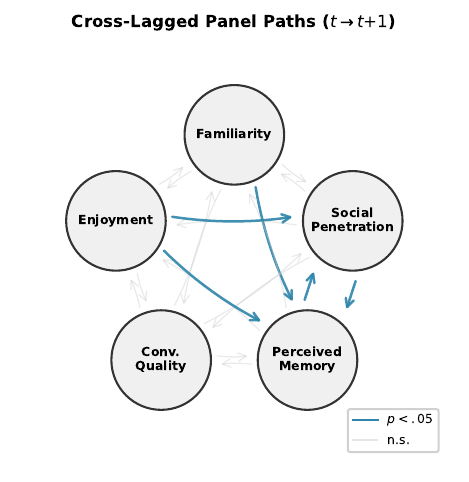}
\Description{Network diagram showing cross-lagged paths from session t to t+1 among five relational constructs, with Perceived Memory receiving significant paths from Enjoyment, Familiarity, and Social Penetration, and sending a significant path to Social Penetration.}
\caption{Cross-lagged paths ($t \to t{+}1$). Perceived Memory emerges as a longitudinal bridge.}
\label{fig:clpm_paths}
\end{subfigure}
\caption{Within-session vs.\ cross-session relational structure among five constructs. Blue/gray lines = significant/non-significant paths ($p < .05$). Full matrices in Appendix~\ref{app:path_coefficients}.}
\label{fig:paths_combined}
\end{figure*}

The delta distributions are not symmetric across all constructs.
Enjoyment exhibits a heavy-tailed, negatively skewed distribution (kurtosis\,=\,3.22, skew\,=\,$-0.89$), indicating that extreme negative changes are more frequent and more severe than extreme positive changes.
This asymmetry motivates treating crashes and surges as distinct phenomena rather than as symmetric tails of a single distribution.

\subsection{Detection and Forecasting Models}
\label{ssec:models}

We frame both crash and surge prediction as binary classification tasks using Elastic-Net Logistic Regression (ENLR).

\paragraph{Task formulation}
We distinguish two prediction tasks:
\begin{itemize}[nosep,leftmargin=*]
  \item \textbf{Detection}: predict whether a crash or surge occurred at session $t$ (i.e., whether $\Delta w_k^{(t)}$ crosses the $\pm 1$\,SD threshold) using features from sessions $1, \ldots, t$---all information available up to and including the session at which the event is observed.
  \item \textbf{Forecasting}: predict whether a crash or surge will occur at session $t$ using features from sessions $1, \ldots, t{-}1$ only---i.e., one-session-ahead prediction before the outcome is observed.
\end{itemize}

\paragraph{Primary configuration and feature-set ablation}
We use ENLR with the full feature set (ENLR\,+\,STP, i.e., Session\,+\,Temporal\,+\,Person-Normalized features) as our \emph{primary} model configuration, reported in the main text.
ENLR's coefficients are directly readable and its regularization structure yields sparse, interpretable solutions; the STP feature set uses all available behavioral information, avoiding post-hoc feature selection.
As a complementary analysis, we evaluate all seven combinations of the three feature families---Session~(S), Temporal~(T), and Person-Normalized~(P)---for both detection and forecasting: S, T, P, ST, SP, TP, and STP.
Table~\ref{tab:ablation_summary} reports ENLR results for each feature-set combination; this ablation identifies which behavioral channels are most informative for each task$\times$event-type combination and confirms that the crash-surge detectability asymmetry is preserved across all seven configurations.
The same modeling pipeline is applied to both crash and surge events for each construct, as well as to systemic crash and surge events.
ENLR employs inverse-frequency class weights during training, upweighting the minority (event) class to prevent the classifier from defaulting to the majority class.

\paragraph{Cross-validation and evaluation}
All models are evaluated under leave-one-participant-out (LOPO) cross-validation with 24 outer folds.
Hyperparameters are tuned via nested LOPO cross-validation within each outer fold.
Feature importance is assessed via bootstrap stability selection~\cite{meinshausen2010stability} ($B{=}100$ resamples per fold).
We report area under the precision-recall curve (AUPRC) as the primary metric, which is more informative than AUROC for imbalanced classification~\cite{saito2015precision}.

%% file: sections/04_results.tex
\section{Results}
\label{sec:results}

\subsection{Temporal Dynamics of Relational Constructs}
\label{subsec:temporal_dynamics}

We first examine the slow-accumulation side of relationship development: which constructs change over repeated interaction, which matter only within a session, and which appear to carry forward.
Across analyses, a consistent dissociation emerges between \emph{immediate} session quality and \emph{durable} cross-session development.

\begin{table}[t]
\centering
\small
\caption{Linear growth slopes ($\beta$ per session) from fixed-effects panel models. Only Social Penetration shows statistically reliable growth over the 10-session study.}
\label{tab:growth_curves}
\begin{tabular}{@{} l r r l @{}}
\toprule
\textbf{Construct} & \textbf{$\beta$} & \textbf{$p$} & \textbf{Sig.} \\
\midrule
Familiarity         & 0.005 & .879 & n.s. \\
Social Penetration  & 0.081 & .003 & ** \\
Perceived Memory    & 0.001 & .975 & n.s. \\
Conv.\ Quality      & 0.040 & .133 & n.s. \\
Enjoyment           & 0.057 & .069 & n.s. \\
\bottomrule
\end{tabular}

\vspace{2pt}
{\footnotesize $**\, p < .01$;\quad n.s.\,=\,not significant.}
\end{table}

\paragraph{Growth trajectories}
The five constructs do not follow a single developmental trajectory (Table~\ref{tab:growth_curves}).
Only Social Penetration increased significantly over sessions (Eq.~\ref{eq:growth}; $\beta = 0.081$, $p = .003$), consistent with the core prediction of Social Penetration Theory~\cite{altman1973social}; the remaining four constructs showed no reliable linear growth (all $p > .06$).

\paragraph{Immediate session quality vs.\ durable cross-session development}
Figure~\ref{fig:paths_combined} summarizes these associations as network diagrams: panel~(a) shows concurrent (same-session) associations among the five constructs, while panel~(b) shows cross-lagged paths from session $t$ to $t{+}1$; blue edges denote significant associations ($p < .05$).
Within sessions, Conversational Quality dominates: it shows the strongest concurrent association with Enjoyment (Eq.~\ref{eq:concurrent}; $\delta = 0.430$, $p < .001$), and high-quality conversation strongly contributes to whether a session feels good in the moment.
Across sessions, however, Conversational Quality does not carry forward: no significant cross-lagged paths emerge from it (all $p > .07$), suggesting that quality must be maintained session by session rather than accumulated.

Perceived Memory shows the opposite pattern.
Rather than dominating within-session affect, it appears to function as a cross-session bridge.
Perceived Memory at session $t{+}1$ was significantly predicted by Enjoyment (Eq.~\ref{eq:clpm}; $\beta_{j \to k} = 0.343$, $p < .001$), Familiarity ($\beta_{j \to k} = 0.290$, $p = .005$), and Social Penetration ($\beta_{j \to k} = 0.268$, $p = .033$) at session $t$.
A fifth significant cross-lagged path, Enjoyment $\to$ Social Penetration ($\beta_{j \to k} = 0.280$, $p = .022$), further indicates that enjoyable sessions are associated with deeper self-disclosure in the next session.
In turn, Perceived Memory at session $t$ predicted greater Social Penetration at session $t{+}1$ ($\beta_{j \to k} = 0.165$, $p = .001$).
Together, these paths suggest that Perceived Memory is richer than a simple readout of system capability.
Because Familiarity, Enjoyment, and Social Penetration all predict later Perceived Memory, perceived continuity appears to be partly \emph{relationally conditioned}: users may notice, interpret, or attribute memory differently depending on the existing state of the relationship.
The Memory $\to$ Social Penetration path remained significant under person-mean centering ($\beta = 0.128$, $p = .002$; Appendix~\ref{app:robustness}).
The reverse path (Social Penetration $\to$ Memory) was directionally consistent but did not reach significance ($\beta = 0.197$, $p = .114$), as expected given the reduced power of the within-person estimator with $N{=}24$.
Memory and Social Penetration thus show tentative reciprocal reinforcement---each predicting the other at the next session---consistent with a process in which perceived continuity invites disclosure, and disclosure provides material that sustains perceived continuity.

\paragraph{Memory $\rightarrow$ Disclosure $\rightarrow$ Enjoyment: a mediated pathway}
Within sessions, Enjoyment is driven by Conv.\ Quality and Social Penetration ($R^2 = 0.847$), but not directly by Memory ($\delta = 0.038$, $p = .452$).
Given that Memory at session $t$ predicts Social Penetration at $t{+}1$, we test whether memory's association with later enjoyment operates through disclosure rather than directly, fitting the lagged mediation model (Eqs.~\ref{eq:med_a}--\ref{eq:med_b}).
The $a$ path (Memory $\to$ Social Penetration; $a = 0.165$, $p = .001$) and $b$ path (Social Penetration $\to$ Enjoyment; $b = 0.497$, $p < .001$) are both significant.
The indirect effect is significant: $a \times b = 0.082$, 95\% CI $[0.029, 0.155]$ (5{,}000 cluster-bootstrap iterations; bias-corrected CI $[0.032, 0.164]$).
The direct effect is near zero ($c' = 0.005$, $p = .943$), indicating \emph{full mediation}.
A parallel mediation model confirms self-disclosure is the primary channel (Social Penetration: 95\% CI $[0.011, 0.089]$; Conv.\ Quality: 95\% CI $[-0.033, 0.127]$, n.s.).
In summary, perceived memory is associated with later enjoyment not through recall display alone, but through its association with deeper self-disclosure in the next session.

\subsection{Relational Turning Points: Crashes and Surges}
\label{subsec:crash_surge}

The preceding analyses characterized gradual relational development.
We now examine abrupt turning points---crashes and surges---as a distinct layer of longitudinal dynamics.
Three asymmetries organize the results: \emph{observability} (which shifts are legible in the moment), \emph{onset} (which shifts are visible in advance versus only when they occur), and \emph{persistence} (which shifts leave durable relational consequences).
Given the modest sample and low event rates, the analytic emphasis is not absolute classification accuracy but whether positive and negative turning points become behaviorally legible on different timescales.

\paragraph{Observability asymmetry: the surge advantage is primarily an in-session effect}
Table~\ref{tab:en_stp_results} reports AUPRC for same-session detection and next-session forecasting under the primary ENLR\,+\,STP configuration.
The clearest pattern appears in detection.
Surge AUPRC exceeds crash AUPRC for all six targets---Familiarity, Social Penetration, Memory, Conversational Quality, Enjoyment, and Systemic events---with a mean advantage of .072 (surge: .215; crash: .143).
Paired bootstrap tests confirm significant surge-over-crash detection advantages for Social Penetration and Conversational Quality; the remaining targets show the same directional pattern without reaching significance individually.
This pattern is consistent across all seven feature-set combinations (Table~\ref{tab:ablation_summary}): surges remain more detectable than crashes.
The reverse test (crash $>$ surge) is not significant for any construct in either detection or forecasting, confirming that the asymmetry is one-directional.

Importantly, this asymmetry weakens in forecasting.
For next-session prediction, the mean difference between surges and crashes shrinks substantially (surge: .181; crash: .170).
Within each event type, the temporal profile is instead reversed: surge detection exceeds forecasting on average (.215 vs.\ .181), whereas crash forecasting exceeds detection (.170 vs.\ .143).
The trained model outperformed simple temporal baselines (previous-session delta, 3-session moving average, marginal event rate) for 10 of 12 targets; for Conversational Quality and systemic crashes, the marginal-rate baseline remained competitive (Appendix~\ref{app:baselines}).
The main asymmetry, then, is not that positive turning points are uniformly easier to predict, but that they are more behaviorally visible once they begin to unfold.

\begin{table}[t]
\centering
\small
\caption{Primary configuration (ENLR\,+\,STP) AUPRC for detection and forecasting. $\dagger$: significant surge $>$ crash in detection; $\ddagger$: significant surge $>$ crash in forecasting ($p < .05$, paired bootstrap). Surge $>$ crash for all 6 detection targets.}
\label{tab:en_stp_results}
\begin{tabular}{@{} l cc cc @{}}
\toprule
 & \multicolumn{2}{c}{\textbf{Detection}} & \multicolumn{2}{c}{\textbf{Forecasting}} \\
\cmidrule(lr){2-3} \cmidrule(lr){4-5}
\textbf{Construct} & Crash & Surge & Crash & Surge \\
\midrule
Familiarity                       & .154 & .219          & .237 & .159 \\
Social Penetr.$^\dagger$          & .073 & \textbf{.197} & .196 & .165 \\
Memory$^\ddagger$                 & .126 & .200          & .119 & \textbf{.169} \\
Conv.\ Quality$^\dagger$          & .204 & \textbf{.300} & .145 & .249 \\
Enjoyment                         & .136 & .185          & .116 & .171 \\
Systemic                          & .164 & .187          & .205 & .174 \\
\midrule
Mean                     & .143 & .215          & .170 & .181 \\
\bottomrule
\end{tabular}
\end{table}

\begin{table}[t]
\centering
\small
\caption{Mean ENLR AUPRC across constructs for each feature-set combination. \textbf{Bold} marks the best feature set within each column. The surge detectability advantage is preserved across all seven configurations.}
\label{tab:ablation_summary}
\begin{tabular}{@{} l cc cc @{}}
\toprule
 & \multicolumn{2}{c}{\textbf{Detection}} & \multicolumn{2}{c}{\textbf{Forecasting}} \\
\cmidrule(lr){2-3} \cmidrule(lr){4-5}
\textbf{Feature Set} & Crash & Surge & Crash & Surge \\
\midrule
S (62)     & \textbf{.145} & .179 & .155 & .192 \\
T (103)    & .115 & .167 & .123 & .219 \\
P (186)    & .127 & .191 & \textbf{.173} & .166 \\
ST (165)   & .123 & .188 & .143 & \textbf{.224} \\
SP (248)   & .142 & .172 & .167 & .184 \\
TP (289)   & .124 & .201 & .158 & .183 \\
STP (351)  & .143 & \textbf{.215} & .170 & .181 \\
\bottomrule
\end{tabular}
\end{table}

\paragraph{Onset asymmetry: crashes split into drift-type and local-failure events}
Crashes show a different temporal profile.
Forecasting slightly exceeds detection on average for crashes (mean .170 vs.\ .143), but this aggregate masks two distinct failure modes.
For Familiarity, Social Penetration, and Systemic events, forecasting yields numerically higher AUPRC than same-session detection, suggesting a \emph{gradual-onset} mode in which deterioration becomes visible first as cross-session drift.
Conversational Quality shows the opposite pattern: detection (0.204) exceeds forecasting (0.145), indicating a more \emph{local} failure mode that is best characterized by within-session cues rather than prior-session trends.

This split aligns with the temporal dynamics in Section~\ref{subsec:temporal_dynamics}.
There, Conversational Quality was the strongest driver of enjoyment within the same session but showed no significant cross-lagged carryover.
Here, its crashes are likewise more detectable than forecastable.
Taken together, the two analyses suggest that Conversational Quality is primarily an interaction-local variable at both the gradual and event levels: it strongly shapes how a session feels, but its failures are largely realized within that session rather than accumulated across sessions.
By contrast, Familiarity and Social Penetration show more cumulative relational structure and correspondingly more forecastable crash profiles.

\paragraph{Different behavioral signatures for positive and negative shifts}
The feature analyses reinforce this distinction.
Crash forecasting depends most strongly on person-normalized prosodic and visual deviations, with stable predictors including valence variability, F0 variability, and backchannel-ratio deviation (Table~\ref{tab:top_features} in Appendix~\ref{app:additional}).
This suggests that some crashes are visible not as a single salient behavioral marker but as deviation from the participant's own baseline.
Surges look different.
Same-session surge detection relies more on cues of interactional expansion and responsiveness---backchannel-ratio variability, loudness, and semantic-alignment change among the most stable predictors.
When surges are forecastable, they draw more heavily on temporal continuity features such as conversational depth, personalized openings, and felt-understanding trends.
Ablation results point in the same direction: crash detection performs best with simpler feature sets, whereas surge detection benefits from richer multimodal and historical context (Table~\ref{tab:ablation_summary}).

\paragraph{Persistence asymmetry: enjoyment surges last longer than enjoyment crashes recover}
Turning points also differ in what happens \emph{after} the event.
For four constructs, crash recovery and surge persistence rates are broadly comparable (53--72\%; Table~\ref{tab:persistence}).
Enjoyment is the notable exception.
Enjoyment surges persist 75\% of the time, whereas enjoyment crashes recover only 52\% ($p < .001$, permutation test; robust across all 24 jackknife folds).
The asymmetry between crashes and surges thus extends beyond behavioral detectability.
At least for enjoyment, positive turning points appear to generate more durable momentum than negative turning points are able to undo.

\begin{table}[t]
\centering
\small
\caption{Crash recovery and surge persistence rates per construct. Recovery: proportion of crashes followed by a return to pre-crash level. Persistence: proportion of surges maintained at or above post-surge level. $^{***}$: $p < .001$, permutation test; robust across all 24 jackknife folds.}
\label{tab:persistence}
\begin{tabular}{@{} l cc cc @{}}
\toprule
 & \multicolumn{2}{c}{\textbf{Crash Recovery}} & \multicolumn{2}{c}{\textbf{Surge Persistence}} \\
\cmidrule(lr){2-3} \cmidrule(lr){4-5}
\textbf{Construct} & $n$ & \% & $n$ & \% \\
\midrule
Familiarity        & 22/35 & 62.9 & 22/34 & 64.7 \\
Social Penetr.     & 13/18 & 72.2 & 23/32 & 71.9 \\
Memory             & 16/24 & 66.7 & 20/29 & 69.0 \\
Conv.\ Quality     & 18/34 & 52.9 & 25/46 & 54.3 \\
Enjoyment$^{***}$  & 11/21 & 52.4 & 21/28 & 75.0 \\
\bottomrule
\end{tabular}
\end{table}

Overall, the event analysis points to an asymmetric temporal ecology of longitudinal human-AI relationships.
Positive shifts are most distinctive in their in-session observability; some negative shifts are distinctive in their advance warning through person-specific drift; and affective surges can be especially durable once they occur.

%% file: sections/05_discussion.tex
\section{Discussion}
\label{sec:discussion}

Taken together, the results argue against a simple view of repeated human-AI interaction as either steady growth or isolated sessions.
Instead, they point to two complementary layers of relational development: a slower cross-session process centered on perceived memory and self-disclosure, and abrupt crash--surge turning points that expose asymmetric intervention windows.
We discuss these in turn: first, what the cross-session dynamics reveal about the role of perceived memory, and second, what the crash--surge asymmetries imply for adaptive intervention.

\subsection{Perceived Memory as a Relational Appraisal}
\label{subsec:memory_backbone}

What made a session enjoyable in the moment was not the same as what appeared to sustain the relationship across sessions.
Session quality mattered primarily within a session, whereas perceived memory was the construct most clearly tied to later relational movement.
Importantly, perceived memory did not behave like a simple readout of system capability.
Because it was shaped by prior familiarity, enjoyment, and self-disclosure, memory appears to function partly as a \emph{relational appraisal}: users feel remembered not only when the agent recalls details, but when the broader interaction already feels coherent and rewarding.

The mediated pathway further clarifies why this matters.
The data are consistent with the idea that memory contributes to later enjoyment mainly by supporting deeper next-session self-disclosure, rather than by directly making the interaction more pleasurable.
This shifts the design target for memory-augmented agents: the goal is not simply to display recall, but to use continuity in ways that reopen prior topics, acknowledge personal context, and invite elaboration---for instance, by referencing a previously shared concern and asking how it has developed---rather than merely demonstrating that the system remembers.

The interviews point to the same mechanism.
One participant described how memory facilitated progressive deepening: \emph{``InteLLA remembered what we had talked about in previous sessions really well, and that became a springboard for conversations to expand---I think that's what made the conversations enjoyable.''}
Conversely, when memory failed, the consequences could cascade beyond a single session: \emph{``Once InteLLA forgot my cat's name and asked the same questions again, ... after that I switched to treating it more as speaking practice rather than trying to build a relationship.''}
This suggests that memory failures do more than lower satisfaction in a single session; they can change the \emph{relational frame} through which the user encounters the system---from relationship-building to utility.
This reframing is also consistent with the view that people hold machines to different standards than they hold humans~\cite{hidalgo2021judge}: a lapse that might be forgiven in a human interlocutor instead prompted the participant to downgrade the agent to a tool, a human-AI-specific deviation from the relational scaffold rather than a straightforward analogue of human--human repair.

\subsection{Asymmetric Intervention Windows}
\label{subsec:asymmetric_intervention}

The second implication is that longitudinal adaptation should be temporally asymmetric.
The crash-surge analyses point to a concrete architectural requirement: adaptive agents need two monitoring systems operating on different timescales.

For crashes, the key distinction is between gradual-onset and sudden failures.
Crashes in familiarity, self-disclosure, and systemic rapport were more forecastable than detectable, suggesting a \emph{drift} mode in which deterioration becomes visible first as cross-session deviation from the user's own baseline.
Session quality shows the opposite pattern: its crashes were more detectable than forecastable, indicating a more \emph{local} failure mode that must be handled within the session itself.
This split aligns with the cross-lagged results, where session quality was the strongest within-session driver of enjoyment but showed no carryover, while familiarity and self-disclosure exhibited more cumulative relational structure.

For surges, the picture is reversed.
Positive turning points were more behaviorally visible in same-session interaction than crashes across all detection targets, and they often benefited from richer multimodal context.
This implies that agents do not necessarily need to predict surges far in advance; instead, they should recognize and capitalize on them in the moment.
Interview data suggest that small acts of personalization can trigger such moments: one participant noted \emph{``even just being called by my name made me feel like it was an extension of yesterday's conversation.''}
Because enjoyment surges persisted reliably (75\%) while enjoyment crashes recovered less often (52\%), reinforcing a positive shift may have durable payoff.

This directional pattern contrasts with the classic expectation that ``bad is stronger than good'' in relationship science~\cite{baumeister2001bad}.
In our data, positive shifts were often more behaviorally traceable than negative ones.
Given that the asymmetry is strongest for a subset of constructs rather than uniform, we view this contrast as suggestive rather than definitive.
The practical takeaway, however, is clear: adaptive human-AI systems should be \emph{asymmetric} in how they intervene---proactively monitoring cross-session drift to prevent gradual crashes, while reactively amplifying surges when they emerge.

\subsection{Limitations}
\label{subsec:limitations}

First, all data were collected with InteLLA, a single memory-augmented LLM agent; the specific dynamics may be architecture-dependent and should not be assumed to transfer to other conversational-agent designs.

Second, absolute detection performance remains modest (AUPRC 0.073--0.300), constituting a proof of concept rather than a deployment-ready system.

Third, on sample size: our $N{=}24$ is small in absolute terms, but is in line with prior longitudinal studies of human-AI/agent relationships, where repeated multi-session designs make large samples costly and small $N$ is standard---e.g., Laban et al.~\cite{laban2024opening} (39 participants, 10 sessions over 5 weeks), Skjuve et al.~\cite{skjuve2022longitudinal} (25 Replika users over 12 weeks), and in-home relational-robot studies with $N{=}13$ over 1--4 months~\cite{yamazaki2023} and $N{=}9$ over 6 weeks~\cite{abendschein2022novelty}.
Crucially, our central claims concern \emph{within-person} change (2{,}270 ratings; 202 transitions) and are stress-tested via permutation tests ($p<.001$), cluster bootstrap, and jackknife (robust in 24/24 folds; Appendix~\ref{app:robustness}), so the small $N$ bounds the \emph{generalizability} of the findings rather than the validity of the dynamics we report. Statistical power nonetheless remains limited for complex models such as the cross-lagged panel: our fixed-effects lagged panel regression absorbs time-invariant between-person differences but remains a short-panel specification rather than a full Random-Intercept CLPM; person-mean centering yields directionally consistent results (Appendix~\ref{app:robustness}), and a formal RI-CLPM decomposition would require a substantially larger sample.

Fourth, generalizability is further constrained by the composition and setting of the sample. Participants were English-proficient university students interacting in English, so the findings are monocultural and language-specific and should not be extrapolated to other cultures, age groups, or demographic populations without further study. Relatedly, a student population accustomed to structured self-disclosure in educational settings may exhibit elevated baseline disclosure that could compress relational formation relative to more typical populations; we note, however, that our mediation results concern within-person change rather than absolute disclosure levels, which partially mitigates this concern.

Fifth, sessions were recorded on participants' own webcams, so device heterogeneity may introduce measurement noise, particularly for gaze-related signals such as focused gaze. Our person-normalized features (participant deviations and $z$-scores) interpret visual cues relative to each participant's own baseline rather than comparing absolute values across participants or devices, which reduces---but does not eliminate---this concern.

Finally, the study captures a limited set of relational constructs. Familiarity, self-disclosure, perceived memory, conversational quality, and enjoyment span core facets of relational development, but complementary dimensions such as trust, emotional attachment, anthropomorphism, and perceived usefulness could yield a more comprehensive account of long-term human-AI relationships and are a natural direction for future work.

%% file: sections/06_conclusion.tex
\section{Conclusion}

Repeated human-AI conversation appears to develop through two coupled mechanisms operating at different timescales.
First, what makes a session satisfying is not the same as what sustains a relationship across sessions: Conversational Quality matters immediately, whereas Perceived Memory functions as a longitudinal bridge---relationally conditioned by prior interaction state rather than reflecting system capability alone---whose link to later enjoyment is carried through self-disclosure.
This positions memory as a relational appraisal that catalyzes deeper conversation, not a mere recall display.
Second, longitudinal human-AI relationships are punctuated by discrete crashes and surges that are partially traceable in multimodal behavior and that open different intervention windows.
The evidence suggests that some crashes are best handled through cross-session drift monitoring, while surges are best recognized and reinforced in the moment.

Together, these findings frame human-AI relational development as both slow accumulation and abrupt turning points.
For design, memory should be built to facilitate disclosure, and adaptive agents should combine proactive crash prevention with reactive surge reinforcement.
To support reproducibility, we will publicly release all analysis code, feature extraction pipelines, and model implementations.

%% file: sections/07_safety_statement.tex
\section{Safe and Responsible Innovation Statement}
\label{sec:safety}

The study protocol was reviewed and approved by the Ethics Review Committee on Research with Human Subjects of Waseda University prior to data collection.
All 24 participants provided informed consent covering the collection of text, audio, and video data across up to 10 sessions, post-session questionnaire responses, and optional semi-structured interviews.
Participants were explicitly informed that their multimodal data would be used for research on human-AI relational dynamics.

%% file: sections/09_acknowledgment.tex
\begin{acks}
This research was supported by the project
"Innovative Information and Communication Technology (Beyond 5G (6G)) Fund Project / Research on an Automatic Evaluation Platform for Highly Reliable Multimodal Conversational AI Agents in the Beyond 5G Era (JPJ012368C-10301)" by the National Institute of Information and Communications Technology (NICT),
and "Adaptable and Seamless Technology transfer Program through Target-driven
R\&D (A-STEP) / Development of a Conversational AI Agent Platform for Diagnostic Assessment and Learning Assistance (JPMJTT24J3)" by Japan Science and Technology Agency (JST).
\end{acks}

%% file: sections/08_appendix.tex

\section{Questionnaire Items and Reliability}
\label{app:questionnaire}

\subsection{Post-Session Questionnaire}
After each of the 10 sessions, participants completed a 10-item questionnaire (7-point Likert scale; 1\,=\,strongly disagree, 7\,=\,strongly agree) capturing five relational constructs (two items each).
The items and their construct mappings are:

\begin{enumerate}[leftmargin=*, nosep]
    \item I feel a sense of familiarity with this conversational AI.\\
    \mbox{}\hfill \emph{Familiarity ($w_1$)}
    \item I feel that this conversational AI empathizes with my feelings.\\
    \mbox{}\hfill \emph{Familiarity ($w_1$)}
    \item I feel comfortable talking to this conversational AI about a wide range of topics regarding various areas of my life.\\
    \mbox{}\hfill \emph{Social Penetration ($w_2$)}
    \item I feel comfortable talking to this conversational AI about deeply personal matters, such as my worries and emotions.\\
    \mbox{}\hfill \emph{Social Penetration ($w_2$)}
    \item I feel that this conversational AI remembers the content of our past conversations.\\
    \mbox{}\hfill \emph{Perceived Memory ($w_3$)}
    \item I feel that this conversational AI understands me.\\
    \mbox{}\hfill \emph{Perceived Memory ($w_3$)}
    \item Conversations with this conversational AI feel natural.\\
    \mbox{}\hfill \emph{Conv.\ Quality ($w_4$)}
    \item This conversational AI's way of speaking is pleasant.\\
    \mbox{}\hfill \emph{Conv.\ Quality ($w_4$)}
    \item It is fun to talk with this conversational AI.\\
    \mbox{}\hfill \emph{Enjoyment ($w_5$)}
    \item I would like to talk to this conversational AI again.\\
    \mbox{}\hfill \emph{Enjoyment ($w_5$)}
\end{enumerate}

\subsection{Reliability}
Table~\ref{tab:reliability} reports per-construct internal consistency.

\begin{table}[H]
\centering
\small
\caption{Internal consistency of session-level self-report constructs.}
\label{tab:reliability}
\begin{tabular}{lccc}
\toprule
\textbf{Construct} & \textbf{Items} & \textbf{Inter-item $r$} & \textbf{$\alpha$} \\
\midrule
Familiarity ($w_1$)      & Q1--Q2  & .752 & .858 \\
Social Penetration ($w_2$) & Q3--Q4  & .768 & .869 \\
Perceived Memory ($w_3$)  & Q5--Q6  & .637 & .778 \\
Conv.\ Quality ($w_4$)    & Q7--Q8  & .739 & .850 \\
Enjoyment ($w_5$)         & Q9--Q10 & .786 & .880 \\
\midrule
Overall (Q1--Q10)         & all     & --   & .942 \\
\bottomrule
\end{tabular}
\end{table}

\section{Psychometric Validation}
\label{app:psychometric}

Confirmatory factor analysis supports the five-factor structure (CFI\,=\,.965, RMSEA\,=\,.107; significantly better than a one-factor model, $\Delta\chi^2(10) = 222.4$, $p < .001$), with all AVE values above .64.
The elevated RMSEA likely reflects the small sample and few items per factor rather than model misspecification, as CFI exceeds the .95 threshold and all factor loadings are strong.
Several session-level HTMT ratios exceed .90 (up to .98 for Memory-Conv.\ Quality), but delta-level HTMT is substantially lower (mean 0.725), and the detectability asymmetry is preserved under a reduced three-factor structure, supporting retention of five constructs.

\paragraph{Delta-level discriminant validity}
The high HTMT ratios reported above (e.g., Memory-Conv.\ Quality = 0.98) reflect between-person convergence in \emph{raw} session-level ratings.
Because our crash-surge analyses operate on session-to-session \emph{deltas}, we recomputed HTMT on delta scores.
Table~\ref{tab:delta_htmt} shows that delta-level HTMT is substantially lower: mean HTMT drops from 0.823 to 0.725, and the number of pairs exceeding the 0.90 threshold drops from four to one.
The two most problematic pairs---Memory-Conv.\ Quality and Familiarity-Conv.\ Quality---drop below 0.80, confirming that the five constructs capture distinct change dynamics even where their absolute levels converge.
One pair (Fam-SocPen) increases from .911 to .959 at the delta level; the three-factor robustness check below confirms that the detectability asymmetry is preserved under a reduced factor structure.

\begin{table}[H]
\centering
\small
\caption{HTMT ratios at the raw (session-level) and delta (session-to-session change) levels. Delta-level HTMT is lower for 6 of 10 pairs, with the largest reductions in the pairs that threaten discriminant validity at the raw level.}
\label{tab:delta_htmt}
\begin{tabular}{@{} l c c c @{}}
\toprule
\textbf{Construct Pair} & \textbf{Raw} & \textbf{Delta} & \textbf{$\Delta$} \\
\midrule
Fam-SocPen     & .911 & .959 & $-$.048 \\
Fam-Memory     & .914 & .613 & +.302 \\
Fam-ConvQual   & .973 & .792 & +.181 \\
Fam-Enjoyment  & .603 & .661 & $-$.058 \\
SocPen-Memory  & .887 & .601 & +.285 \\
SocPen-ConvQual & .840 & .730 & +.110 \\
SocPen-Enjoy   & .554 & .780 & $-$.225 \\
Mem-ConvQual   & .980 & .746 & +.235 \\
Mem-Enjoyment  & .772 & .513 & +.259 \\
ConvQual-Enjoy & .800 & .853 & $-$.053 \\
\midrule
Mean            & .823 & .725 & +.099 \\
Pairs $>$ .90   & 4    & 1    & \\
\bottomrule
\end{tabular}
\end{table}

\paragraph{Three-factor robustness}
We tested whether the detectability asymmetry holds under a reduced three-factor structure (Impression = Familiarity + Conv.\ Quality; Depth = Social Penetration + Enjoyment; Memory unchanged).
We ran the full LOPO detection pipeline on the three merged constructs: surges are more detectable than crashes for all three (Table~\ref{tab:three_factor_robust}).

\begin{table}[H]
\centering
\small
\caption{Detectability asymmetry under the 5-factor and reduced 3-factor structures. The surge detectability advantage is preserved.}
\label{tab:three_factor_robust}
\begin{tabular}{@{} l c c @{}}
\toprule
\textbf{Dimension} & \textbf{5-Factor} & \textbf{3-Factor} \\
\midrule
Detectability (surge $>$ crash) & 5/5 constructs & 3/3 constructs \\
\bottomrule
\end{tabular}
\end{table}

\section{Threshold Sensitivity Analysis}
\label{app:sensitivity}

To verify that the crash-surge asymmetry is not an artifact of the 1-SD threshold choice, we repeated the core analyses at 0.75, 1.0, and 1.25 SD thresholds.
Table~\ref{tab:sensitivity} reports crash detection AUPRC across thresholds.
The key patterns are preserved: detection performance remains in a comparable range at 0.75 and 1.0 SD before declining at 1.25 SD due to reduced event counts (e.g., Familiarity drops to 14 crashes).

\begin{table}[H]
\centering
\small
\caption{Threshold sensitivity analysis for crash detection (ENLR). Detection AUPRC is shown across SD thresholds. Detection performance is broadly comparable at 0.75 and 1.0~SD before declining at 1.25~SD due to reduced event counts.}
\label{tab:sensitivity}
\begin{tabular}{@{} l c c c @{}}
\toprule
& \textbf{0.75 SD} & \textbf{1.0 SD} & \textbf{1.25 SD} \\
\midrule
\multicolumn{4}{@{}l}{\emph{Crash detection AUPRC}} \\
\addlinespace
Familiarity      & 0.171 & 0.171 & 0.099 \\
Social Penetr.   & 0.080 & 0.080 & 0.080 \\
Memory           & 0.297 & 0.096 & 0.061 \\
Conv.\ Quality   & 0.155 & 0.122 & 0.071 \\
Enjoyment        & 0.073 & 0.073 & 0.073 \\
\bottomrule
\end{tabular}

\vspace{2pt}
{\footnotesize For Social Penetration and Enjoyment, the three SD thresholds produce nearly identical event sets, yielding stable AUPRC values across thresholds.}
\end{table}

\section{Robustness Analyses}
\label{app:robustness}

We report three robustness analyses addressing sample size ($N{=}24$) limitations. Discriminant validity analyses (delta-level HTMT and three-factor robustness) are reported in Appendix~\ref{app:psychometric}.

\paragraph{Person-mean centering check}
The main lagged regressions already include participant fixed effects, which absorb time-invariant between-person differences. As an additional within-person parameterization, we re-estimated the lagged models using participant-mean-centered variables. Under person-mean centering, the Memory$\rightarrow$Disclosure path remains significant ($\beta = 0.128$, $p = .002$), consistent with the fixed-effects estimate and providing further reassurance that the pattern is not purely driven by between-person differences. A full Random-Intercept CLPM would formally decompose stable trait variance from within-person dynamics but requires a substantially larger sample than our $N{=}24$.

\paragraph{Permutation test}
To test whether the observed persistence asymmetry could arise by chance under the null hypothesis of crash-surge symmetry, we ran a permutation test ($B{=}2{,}000$) that randomly swaps crash and surge labels within each construct and transition.
The persistence asymmetry is highly significant ($p < .001$): the observed difference (surge persistence $-$ crash recovery = 0.246) exceeds all 2,000 permuted values.

\paragraph{Jackknife sensitivity}
We dropped each of the 24 participants in turn and recomputed the persistence asymmetry.
The persistence asymmetry (surge persistence $>$ crash recovery) holds under \emph{every} jackknife fold (24/24).
No single participant's removal changes the finding, confirming that the result is not driven by individual outliers.

\FloatBarrier
\section{Path Coefficient Matrices}
\label{app:path_coefficients}

Tables~\ref{tab:clpm_matrix} and~\ref{tab:concurrent_matrix} report the full coefficient matrices for the cross-lagged and concurrent path analyses visualized in Figure~\ref{fig:paths_combined}.

\begin{table*}[!ht]
\centering
\small
\caption{Cross-lagged panel model coefficients. Each cell shows the standardized coefficient $\beta$ for the path Predictor$(t) \to$ Outcome$(t{+}1)$, controlling for the autoregressive term and session. Diagonal entries (gray) are autoregressive stability coefficients. Cluster-robust SEs grouped by participant ($N = 24$, 202 transitions).}
\label{tab:clpm_matrix}
\begin{tabular}{@{} l c c c c c @{}}
\toprule
& \multicolumn{5}{c}{\textbf{Outcome ($t{+}1$)}} \\
\cmidrule(lr){2-6}
\textbf{Predictor ($t$)} & Familiarity & Social Penetr. & Perc.\ Memory & Conv.\ Quality & Enjoyment \\
\midrule
Familiarity        & \cellcolor{gray!15}.213$^{**}$  & .144            & \textbf{.290}$^{**}$   & .103            & .001           \\
Social Penetration & $-$.094        & \cellcolor{gray!15}.237$^{***}$ & \textbf{.268}$^{*}$    & .090            & .063           \\
Perceived Memory   & $-$.017        & \textbf{.165}$^{**}$  & \cellcolor{gray!15}.090          & .087            & .069           \\
Conv.\ Quality     & .061           & .170            & .233            & \cellcolor{gray!15}.113          & $-$.032        \\
Enjoyment          & .056           & \textbf{.280}$^{*}$   & \textbf{.343}$^{***}$  & .305            & \cellcolor{gray!15}.360$^{***}$ \\
\bottomrule
\end{tabular}

\vspace{2pt}
{\footnotesize Bold = significant cross-lagged path ($p < .05$). $^{*}p<.05$, $^{**}p<.01$, $^{***}p<.001$.}
\end{table*}

\begin{table*}[!ht]
\centering
\small
\caption{Concurrent (same-session) regression coefficients. Each column is a separate regression of the outcome on the other four constructs, controlling for session and participant fixed effects. Cluster-robust SEs grouped by participant ($N = 24$, 227 observations).}
\label{tab:concurrent_matrix}
\begin{tabular}{@{} l c c c c c @{}}
\toprule
& \multicolumn{5}{c}{\textbf{Outcome}} \\
\cmidrule(lr){2-6}
\textbf{Predictor} & Familiarity & Social Penetr. & Perc.\ Memory & Conv.\ Quality & Enjoyment \\
& ($R^2{=}.820$) & ($R^2{=}.870$) & ($R^2{=}.761$) & ($R^2{=}.862$) & ($R^2{=}.847$) \\
\midrule
Familiarity        & ---            & \textbf{.203}$^{*}$   & .157            & \textbf{.304}$^{***}$ & .005           \\
Social Penetration & \textbf{.270}$^{**}$  & ---            & \textbf{.298}$^{*}$    & .033            & \textbf{.276}$^{***}$ \\
Perceived Memory   & .094           & \textbf{.134}$^{*}$   & ---             & \textbf{.192}$^{**}$  & .038           \\
Conv.\ Quality     & \textbf{.392}$^{***}$ & .032           & \textbf{.412}$^{**}$   & ---             & \textbf{.430}$^{***}$ \\
Enjoyment          & .006           & \textbf{.266}$^{***}$ & .083            & \textbf{.430}$^{***}$ & ---            \\
\bottomrule
\end{tabular}

\vspace{2pt}
{\footnotesize Bold = significant ($p < .05$). $^{*}p<.05$, $^{**}p<.01$, $^{***}p<.001$.}
\end{table*}

\FloatBarrier
\section{Additional Tables and Figures}
\label{app:additional}

This section provides the top stable features and the complete feature inventories referenced in Section~\ref{ssec:features}: Table~\ref{tab:top_features} lists the top 5 most stable features per task and event type, Table~\ref{tab:text_features} lists the 42 session-level text features organized by conceptual family, and Table~\ref{tab:audio_features} lists the 16 session-level audio features.

\begin{table}[H]
\centering
\small
\caption{Top 5 most stable features per task and event type, selected from the best-performing feature-set configuration per construct. All listed features were selected in $\geq$22 of 24 LOPO folds within their strongest construct. Feature type: S\,=\,session-level, T\,=\,temporal, P\,=\,person-normalized. Modality: Txt\,=\,text, Aud\,=\,audio, Vid\,=\,video.}
\label{tab:top_features}
\resizebox{\columnwidth}{!}{%
\begin{tabular}{@{} l l c l l c @{}}
\toprule
& \textbf{Crash Detection} & \textbf{Mod.} & & \textbf{Surge Detection} & \textbf{Mod.} \\
\midrule
1 & Vulnerability level var.\ (P) & Txt & 1 & Backchannel ratio var.\ (P) & Aud \\
2 & Question about bot (S) & Txt & 2 & Loudness mean var.\ (P) & Aud \\
3 & Focused gaze \% (S) & Vid & 3 & F0 slope var.\ (P) & Aud \\
4 & Bot question dominance (S) & Txt & 4 & Vulnerability level var.\ (P) & Txt \\
5 & Semantic alignment (T) & Txt & 5 & Semantic alignment $\Delta$ (T) & Txt \\
\midrule
& \textbf{Crash Forecasting} & & & \textbf{Surge Forecasting} & \\
\midrule
1 & Valence max var.\ (P) & Vid & 1 & Conversational depth (S) & Txt \\
2 & Backchannel ratio (S) & Aud & 2 & Personalized opening (S) & Txt \\
3 & F0 std var.\ (P) & Aud & 3 & Felt understanding trend (T) & Txt \\
4 & Backchannel ratio dev.\ (P) & Aud & 4 & Backchannel ratio $\Delta$ (T) & Aud \\
5 & Question about bot (S) & Txt & 5 & Emotional attunement $\Delta$ (T) & Txt \\
\bottomrule
\end{tabular}}
\end{table}

\begin{table*}[!ht]
\centering
\small
\caption{Text feature inventory (42 session-level features). Features are organized by conceptual family.}
\label{tab:text_features}
\begin{tabular}{@{} l l p{12cm} @{}}
\toprule
\textbf{Family} & \textbf{Count} & \textbf{Features} \\
\midrule
Emotional responsiveness & 6 & Emotion expression rate, empathy markers, sentiment polarity, emotional reciprocity, valence shifts, affect intensity \\
Disclosure dynamics & 5 & Vulnerability level (mean, max), self-disclosure markers, disclosure depth, disclosure breadth \\
Conversational balance & 5 & Turn length ratio, question-answer balance, topic initiation ratio, interruption rate, conversational dominance \\
Memory and continuity & 6 & Memory references, prior-session callbacks, name usage count, repetition score, continuity markers, shared history references \\
Answer quality & 4 & Informativeness, relevance, coherence, specificity \\
Opening/closing patterns & 4 & Greeting elaboration, closing warmth, session-bridging references, farewell sentiment \\
Semantic alignment & 4 & User-bot semantic similarity, topic overlap, vocabulary convergence, style matching \\
Topic novelty & 4 & Topic novelty score, topic diversity, new-topic ratio, topic depth \\
LLM-derived semantic & 4 & Conversational acts distribution, pragmatic appropriateness, question about bot, actionability markers \\
\bottomrule
\end{tabular}
\end{table*}

\begin{table}[H]
\centering
\small
\caption{Audio feature inventory (16 session-level features).}
\label{tab:audio_features}
\begin{tabular}{@{} l l @{}}
\toprule
\textbf{Category} & \textbf{Features} \\
\midrule
Prosody & F0 mean, F0 std, F0 slope, loudness mean \\
 & jitter, shimmer, HNR, MFCC1 \\
Turn-taking & Speech time ratio, backchannel ratio \\
 & Response gap, overlap rate \\
Emotion dynamics & Arousal mean, arousal max \\
 & Valence mean, valence std \\
\bottomrule
\end{tabular}
\end{table}

\FloatBarrier
\section{LLM Annotation Validation}
\label{app:annotation_validation}

To assess the reliability of GPT-4.1-derived text annotations, two human annotators independently annotated a stratified random sample of 300 utterances (150 user, 150 bot) plus 100 turn pairs.
Table~\ref{tab:iaa_binary} reports inter-annotator agreement for binary fields, and Table~\ref{tab:iaa_ordinal} for ordinal fields.

Several low-prevalence binary fields (memory test, memory reference, praise bot, offer support) exhibit near-zero kappa despite $>$93\% agreement, illustrating the well-known kappa-prevalence paradox~\cite{feinstein1990high}: when nearly all items fall into one category, even small disagreements produce low kappa.
Excluding these fields, mean pairwise Cohen's $\kappa = 0.62$ across the remaining 8 binary annotations, indicating substantial agreement.

Among non-prevalence-affected fields, open ended followup shows poor agreement ($\kappa = 0.34$, 46\% agreement), likely reflecting genuine ambiguity in distinguishing open-ended follow-ups from topic-continuing questions.
For ordinal fields (Table~\ref{tab:iaa_ordinal}), hedge strength ($\alpha = 0.15$) and self disclosure level ($\kappa_w = 0.31$) show limited agreement.
These annotations are intermediate turn-level labels that are aggregated into session-level features; individual annotation noise is partially attenuated through this aggregation, though it remains a source of measurement error.
Among the top stable predictors (Table~\ref{tab:top_features}), vulnerability level variability appears prominently; this feature is derived from vuln level ($\bar{\kappa}_w = 0.73$, $\alpha = 0.75$), which shows substantial agreement.
Features derived from the lowest-agreement annotations---deep followup rate (from open ended followup, $\kappa = 0.34$) and hedging rate (from hedge strength, $\alpha = 0.15$)---do not rank among the top features for any condition.
The most stable predictors---disclosure variability, backchannel ratio variability, and question-about-bot ($\kappa = 0.65$)---are derived from annotations with adequate agreement or from non-LLM sources (audio and video features).
Moreover, because all models are evaluated under LOPO cross-validation, annotation noise can reduce detection performance but cannot inflate it; the reported AUPRC values are therefore conservative with respect to annotation error.

\begin{table}[H]
\centering
\small
\caption{Inter-annotator agreement for binary annotation fields between two human annotators ($n_{\text{user}}=150$, $n_{\text{bot}}=150$, $n_{\text{pair}}=100$). $\kappa$: mean pairwise Cohen's $\kappa$; \%Agr: mean percent agreement. $\dagger$: low-prevalence field (kappa-prevalence paradox).}
\label{tab:iaa_binary}
\begin{tabular}{@{} l l c c @{}}
\toprule
\textbf{Level} & \textbf{Field} & $\kappa$ & \textbf{\%Agr} \\
\midrule
\multirow{5}{*}{User} & emotion present & .61 & 86.2 \\
 & praise bot$^\dagger$ & .13 & 98.2 \\
 & question about bot & .65 & 98.2 \\
 & memory test$^\dagger$ & .00 & 99.6 \\
 & memory frustration$^\dagger$ & .22 & 99.1 \\
\addlinespace
\multirow{4}{*}{Bot} & is question & .88 & 95.1 \\
 & informative & .58 & 91.1 \\
 & actionable & .58 & 97.8 \\
 & memory reference$^\dagger$ & .00 & 98.2 \\
\addlinespace
\multirow{5}{*}{Pair} & ack emotion & .53 & 76.0 \\
 & validate emotion & .44 & 73.3 \\
 & offer support$^\dagger$ & .05 & 93.3 \\
 & open ended followup & .34 & 46.0 \\
 & topic shift & .67 & 94.7 \\
\bottomrule
\end{tabular}
\end{table}

\begin{table}[H]
\centering
\small
\caption{Inter-annotator agreement for ordinal annotation fields. $\bar{\kappa}_w$: mean pairwise quadratic-weighted Cohen's $\kappa$; $\alpha$: Krippendorff's $\alpha$ (ordinal distance).}
\label{tab:iaa_ordinal}
\begin{tabular}{@{} l l c c @{}}
\toprule
\textbf{Level} & \textbf{Field} & $\bar{\kappa}_w$ & $\alpha$ \\
\midrule
\multirow{2}{*}{User} & emotion valence ($-$1/0/$+$1) & .64 & .64 \\
 & vuln level (0--3) & .73 & .75 \\
\addlinespace
\multirow{2}{*}{Bot} & self disclosure level (0--3) & .31 & .46 \\
 & hedge strength (0--2) & .21 & .15 \\
\bottomrule
\end{tabular}
\end{table}

\FloatBarrier
\section{Simple Baseline Comparison}
\label{app:baselines}

Table~\ref{tab:simple_baselines} compares the trained ENLR model against three simple temporal baselines for forecasting.

\begin{table}[H]
\centering
\small
\caption{Forecasting AUPRC for three simple temporal baselines vs.\ trained ENLR model (best across feature sets) under LOPO cross-validation. Prev.~$\Delta$: previous-session delta sign; 3-MA: 3-session moving-average trend; Marginal: training-set event rate.}
\label{tab:simple_baselines}
\resizebox{\columnwidth}{!}{%
\begin{tabular}{@{} l l r r r r @{}}
\toprule
 & & \multicolumn{3}{c}{\textbf{Simple Baselines}} & \\
\cmidrule(lr){3-5}
\textbf{Construct} & \textbf{Event} & \textbf{Prev.\,$\Delta$} & \textbf{3-MA} & \textbf{Marginal} & \textbf{Trained} \\
\midrule
Familiarity & Crash & .183 & .166 & .117 & \textbf{.281} \\
Social Penetr. & Crash & .059 & .061 & .063 & \textbf{.196} \\
Perc.\ Memory & Crash & .080 & .094 & .079 & \textbf{.145} \\
Conv.\ Quality & Crash & .090 & .088 & \textbf{.181} & .168 \\
Enjoyment & Crash & .107 & .112 & .071 & \textbf{.194} \\
Systemic & Crash & .173 & .126 & .\textbf{.229} & .225 \\
\addlinespace
Familiarity & Surge & .113 & .129 & .118 & \textbf{.301} \\
Social Penetr. & Surge & .135 & .118 & .112 & \textbf{.201} \\
Perc.\ Memory & Surge & .089 & .087 & .099 & \textbf{.259} \\
Conv.\ Quality & Surge & .189 & .134 & .259 & \textbf{.275} \\
Enjoyment & Surge & .159 & .122 & .095 & \textbf{.195} \\
Systemic & Surge & .177 & .158 & .250 & \textbf{.265} \\
\bottomrule
\end{tabular}}
\end{table}

\FloatBarrier
\section{InteLLA System Prompt and Memory Prompts}
\label{app:agent_system_prompt}

\subsection{System Prompt}

The conversational agent uses the following system prompt:

\begin{quote}
\small
\ttfamily\raggedright
You are InteLLA, a friendly, kind, and understanding 24-year-old chatbot. The conversation is with university students.\\
- Use clear, simple English with short sentences and common vocabulary.\\
- Speak naturally but avoid idioms or slang that might confuse learners.\\
- Ignore bad grammar and spelling.

GOAL: Have a smooth, friendly chit-chat conversation for about 5 minutes, \textasciitilde 20--25 turns.

CHIT-CHAT GUIDELINES:\\
- Keep the tone warm, relaxed, and natural---like friendly small talk.\\
- Ask interesting, varied questions that encourage sharing.\\
- Use natural follow-ups connected to the user's last message.\\
- Adapt your tone to the user's mood (friendly, curious, gentle).\\
- Keep acknowledgements short (under 12 words), ending with a comma.\\
- Never repeat or summarize what the user says.\\
- Ask one question at a time---no double questions.\\
- Use short, natural replies (under 20 words, max two sentences).

CLOSING: End with a natural, positive closing (e.g., ``That was a fun talk. I hope you enjoyed our chat!'').
\end{quote}

For sessions after the first, the prompt is appended with the current date, session number, and the summary of previous conversations. When RAG is enabled, retrieved memory context is also injected (see below).

\subsection{Memory Prompts}

\paragraph{Ice-breaker generation} Before each session (except the first), the agent generates a personalized opening:

\begin{quote}
\small
\ttfamily\raggedright
You're reconnecting with a friend you've chatted with a few times. Use these key moments from your last conversation: \{history\}. Write a casual, friendly ice breaker (under 20 words) that naturally refers to the history when possible. If the history lacks useful details, write a generic casual opener instead.
\end{quote}

\paragraph{Session summary update} After each session, the user profile is updated:

\begin{quote}
\small
\ttfamily\raggedright
You are given a previous summary of the conversation and a new conversation history. Your task is to generate an updated summary in no longer than 150 words. If the new conversation adds substantial new information, update the summary. If there is little new information, keep the updated summary very similar to the previous one.
\end{quote}

\paragraph{RAG memory injection} Retrieved memories are injected into the system prompt:

\begin{quote}
\small
\ttfamily\raggedright
Conversation Memory (retrieved context): The following notes come from past sessions or the user's profile. Use them to make the conversation more natural and personal.
\end{quote}

%% file: main.bbl

\begin{thebibliography}{35}


\ifx \showCODEN    \undefined \def \showCODEN     #1{\unskip}     \fi
\ifx \showISBNx    \undefined \def \showISBNx     #1{\unskip}     \fi
\ifx \showISBNxiii \undefined \def \showISBNxiii  #1{\unskip}     \fi
\ifx \showISSN     \undefined \def \showISSN      #1{\unskip}     \fi
\ifx \showLCCN     \undefined \def \showLCCN      #1{\unskip}     \fi
\ifx \shownote     \undefined \def \shownote      #1{#1}          \fi
\ifx \showarticletitle \undefined \def \showarticletitle #1{#1}   \fi
\ifx \showURL      \undefined \def \showURL       {\relax}        \fi
\providecommand\bibfield[2]{#2}
\providecommand\bibinfo[2]{#2}
\providecommand\natexlab[1]{#1}
\providecommand\showeprint[2][]{arXiv:#2}

\bibitem[Abendschein et~al\mbox{.}(2022)]%
        {abendschein2022novelty}
\bibfield{author}{\bibinfo{person}{Bryan Abendschein}, \bibinfo{person}{Autumn Edwards}, {and} \bibinfo{person}{Chad Edwards}.} \bibinfo{year}{2022}\natexlab{}.
\newblock \showarticletitle{Novelty Experience in Prolonged Interaction: A Qualitative Study of Socially-Isolated College Students' In-Home Use of a Robot Companion Animal}.
\newblock \bibinfo{journal}{\emph{Frontiers in Robotics and AI}}  \bibinfo{volume}{9} (\bibinfo{year}{2022}), \bibinfo{pages}{733078}.
\newblock
\href{https://doi.org/10.3389/frobt.2022.733078}{doi:\nolinkurl{10.3389/frobt.2022.733078}}


\bibitem[Altman and Taylor(1973)]%
        {altman1973social}
\bibfield{author}{\bibinfo{person}{Irwin Altman} {and} \bibinfo{person}{Dalmas~A. Taylor}.} \bibinfo{year}{1973}\natexlab{}.
\newblock \bibinfo{booktitle}{\emph{Social Penetration: The Development of Interpersonal Relationships}}.
\newblock \bibinfo{publisher}{Holt, Rinehart \& Winston}.
\newblock


\bibitem[Baltru{\v{s}}aitis et~al\mbox{.}(2018)]%
        {baltrusaitis2018openface}
\bibfield{author}{\bibinfo{person}{Tadas Baltru{\v{s}}aitis}, \bibinfo{person}{Amir Zadeh}, \bibinfo{person}{Yao~Chong Lim}, {and} \bibinfo{person}{Louis-Philippe Morency}.} \bibinfo{year}{2018}\natexlab{}.
\newblock \showarticletitle{{OpenFace 2.0}: Facial Behavior Analysis Toolkit}. In \bibinfo{booktitle}{\emph{Proceedings of the 13th IEEE International Conference on Automatic Face \& Gesture Recognition}}. \bibinfo{pages}{59--66}.
\newblock


\bibitem[Baumeister et~al\mbox{.}(2001)]%
        {baumeister2001bad}
\bibfield{author}{\bibinfo{person}{Roy~F. Baumeister}, \bibinfo{person}{Ellen Bratslavsky}, \bibinfo{person}{Catrin Finkenauer}, {and} \bibinfo{person}{Kathleen~D. Vohs}.} \bibinfo{year}{2001}\natexlab{}.
\newblock \showarticletitle{Bad Is Stronger Than Good}.
\newblock \bibinfo{journal}{\emph{Review of General Psychology}} \bibinfo{volume}{5}, \bibinfo{number}{4} (\bibinfo{year}{2001}), \bibinfo{pages}{323--370}.
\newblock


\bibitem[Berger and Calabrese(1975)]%
        {berger1975uncertainty}
\bibfield{author}{\bibinfo{person}{Charles~R. Berger} {and} \bibinfo{person}{Richard~J. Calabrese}.} \bibinfo{year}{1975}\natexlab{}.
\newblock \showarticletitle{Some Explorations in Initial Interaction and Beyond: Toward a Developmental Theory of Interpersonal Communication}.
\newblock \bibinfo{journal}{\emph{Human Communication Research}} \bibinfo{volume}{1}, \bibinfo{number}{2} (\bibinfo{year}{1975}), \bibinfo{pages}{99--112}.
\newblock


\bibitem[Bohus and Horvitz(2009)]%
        {bohus2009models}
\bibfield{author}{\bibinfo{person}{Dan Bohus} {and} \bibinfo{person}{Eric Horvitz}.} \bibinfo{year}{2009}\natexlab{}.
\newblock \showarticletitle{Models for Multiparty Engagement in Open-World Dialog}. In \bibinfo{booktitle}{\emph{Proceedings of SIGdial}}. \bibinfo{pages}{225--234}.
\newblock


\bibitem[Brandtzaeg et~al\mbox{.}(2022)]%
        {brandtzaeg2022my}
\bibfield{author}{\bibinfo{person}{Petter~Bae Brandtzaeg}, \bibinfo{person}{Marita Skjuve}, {and} \bibinfo{person}{Asbj{\o}rn F{\o}lstad}.} \bibinfo{year}{2022}\natexlab{}.
\newblock \showarticletitle{My AI Friend: How Users of a Social Chatbot Understand Their Human--AI Friendship}.
\newblock \bibinfo{journal}{\emph{Human Communication Research}} \bibinfo{volume}{48}, \bibinfo{number}{3} (\bibinfo{year}{2022}), \bibinfo{pages}{404--429}.
\newblock


\bibitem[Eyben et~al\mbox{.}(2016)]%
        {eyben2016geneva}
\bibfield{author}{\bibinfo{person}{Florian Eyben}, \bibinfo{person}{Klaus~R. Scherer}, \bibinfo{person}{Bj{\"o}rn~W. Schuller}, {et~al\mbox{.}}} \bibinfo{year}{2016}\natexlab{}.
\newblock \showarticletitle{The Geneva Minimalistic Acoustic Parameter Set ({GeMAPS}) for Voice Research and Affective Computing}.
\newblock \bibinfo{journal}{\emph{IEEE Transactions on Affective Computing}} \bibinfo{volume}{7}, \bibinfo{number}{2} (\bibinfo{year}{2016}), \bibinfo{pages}{190--202}.
\newblock


\bibitem[Eyben et~al\mbox{.}(2010)]%
        {eyben2010opensmile}
\bibfield{author}{\bibinfo{person}{Florian Eyben}, \bibinfo{person}{Martin W{\"o}llmer}, {and} \bibinfo{person}{Bj{\"o}rn Schuller}.} \bibinfo{year}{2010}\natexlab{}.
\newblock \showarticletitle{{openSMILE} -- The {M}unich Versatile and Fast Open-Source Audio Feature Extractor}. In \bibinfo{booktitle}{\emph{Proceedings of the 18th ACM International Conference on Multimedia}}. \bibinfo{pages}{1459--1462}.
\newblock


\bibitem[Feinstein and Cicchetti(1990)]%
        {feinstein1990high}
\bibfield{author}{\bibinfo{person}{Alvan~R. Feinstein} {and} \bibinfo{person}{Domenic~V. Cicchetti}.} \bibinfo{year}{1990}\natexlab{}.
\newblock \showarticletitle{High agreement but low kappa: {I}. The problems of two paradoxes}.
\newblock \bibinfo{journal}{\emph{Journal of Clinical Epidemiology}} \bibinfo{volume}{43}, \bibinfo{number}{6} (\bibinfo{year}{1990}), \bibinfo{pages}{543--549}.
\newblock


\bibitem[Gottman(1994)]%
        {gottman1994cascade}
\bibfield{author}{\bibinfo{person}{John~M. Gottman}.} \bibinfo{year}{1994}\natexlab{}.
\newblock \bibinfo{booktitle}{\emph{What Predicts Divorce? The Relationship Between Marital Processes and Marital Outcomes}}.
\newblock \bibinfo{publisher}{Lawrence Erlbaum Associates}.
\newblock


\bibitem[Hidalgo et~al\mbox{.}(2021)]%
        {hidalgo2021judge}
\bibfield{author}{\bibinfo{person}{C{\'e}sar~A. Hidalgo}, \bibinfo{person}{Diana Orghian}, \bibinfo{person}{Jordi Albo-Canals}, \bibinfo{person}{Filipa de Almeida}, {and} \bibinfo{person}{Natalia Martin}.} \bibinfo{year}{2021}\natexlab{}.
\newblock \bibinfo{booktitle}{\emph{How Humans Judge Machines}}.
\newblock \bibinfo{publisher}{MIT Press}, \bibinfo{address}{Cambridge, MA}.
\newblock
\urldef\tempurl%
\url{https://www.judgingmachines.com}
\showURL{%
\tempurl}


\bibitem[Higashinaka et~al\mbox{.}(2016)]%
        {higashinaka2016dialogue}
\bibfield{author}{\bibinfo{person}{Ryuichiro Higashinaka}, \bibinfo{person}{Kotaro Funakoshi}, \bibinfo{person}{Yuka Kobayashi}, {and} \bibinfo{person}{Michimasa Inaba}.} \bibinfo{year}{2016}\natexlab{}.
\newblock \showarticletitle{The Dialogue Breakdown Detection Challenge: Task Description, Datasets, and Evaluation Metrics}. In \bibinfo{booktitle}{\emph{Proceedings of LREC}}. \bibinfo{pages}{3146--3150}.
\newblock


\bibitem[Ho et~al\mbox{.}(2018)]%
        {ho2018psychological}
\bibfield{author}{\bibinfo{person}{Annabell Ho}, \bibinfo{person}{Jeff Hancock}, {and} \bibinfo{person}{Adam~S. Miner}.} \bibinfo{year}{2018}\natexlab{}.
\newblock \showarticletitle{Psychological, Relational, and Emotional Effects of Self-Disclosure After Conversations With a Chatbot}.
\newblock \bibinfo{journal}{\emph{Journal of Communication}} \bibinfo{volume}{68}, \bibinfo{number}{4} (\bibinfo{year}{2018}), \bibinfo{pages}{712--733}.
\newblock


\bibitem[Jo et~al\mbox{.}(2024)]%
        {jo2024longtermmemory}
\bibfield{author}{\bibinfo{person}{Eunkyung Jo}, \bibinfo{person}{Yuin Jeong}, \bibinfo{person}{SoHyun Park}, \bibinfo{person}{Daniel~A. Epstein}, {and} \bibinfo{person}{Young-Ho Kim}.} \bibinfo{year}{2024}\natexlab{}.
\newblock \showarticletitle{Understanding the Impact of Long-Term Memory on Self-Disclosure with Large Language Model-Driven Chatbots for Public Health Intervention}. In \bibinfo{booktitle}{\emph{Proceedings of the 2024 CHI Conference on Human Factors in Computing Systems}}. \bibinfo{publisher}{ACM}.
\newblock


\bibitem[Kanda et~al\mbox{.}(2007)]%
        {kanda2007two}
\bibfield{author}{\bibinfo{person}{Takayuki Kanda}, \bibinfo{person}{Rumi Sato}, \bibinfo{person}{Naoki Saiwaki}, {and} \bibinfo{person}{Hiroshi Ishiguro}.} \bibinfo{year}{2007}\natexlab{}.
\newblock \showarticletitle{A Two-Month Field Trial in an Elementary School for Long-Term Human--Robot Interaction}.
\newblock \bibinfo{journal}{\emph{IEEE Transactions on Robotics}} \bibinfo{volume}{23}, \bibinfo{number}{5} (\bibinfo{year}{2007}), \bibinfo{pages}{962--971}.
\newblock


\bibitem[Laban et~al\mbox{.}(2024a)]%
        {laban2023longterm}
\bibfield{author}{\bibinfo{person}{Guy Laban}, \bibinfo{person}{Arvid Kappas}, \bibinfo{person}{Val Morrison}, {and} \bibinfo{person}{Emily~S. Cross}.} \bibinfo{year}{2024}\natexlab{a}.
\newblock \showarticletitle{Building Long-Term Human--Robot Relationships: Examining Disclosure, Perception and Well-Being Across Time}.
\newblock \bibinfo{journal}{\emph{International Journal of Social Robotics}} \bibinfo{volume}{16}, \bibinfo{number}{5} (\bibinfo{year}{2024}), \bibinfo{pages}{953--979}.
\newblock


\bibitem[Laban et~al\mbox{.}(2024b)]%
        {laban2024opening}
\bibfield{author}{\bibinfo{person}{Guy Laban}, \bibinfo{person}{Arvid Kappas}, \bibinfo{person}{Val Morrison}, {and} \bibinfo{person}{Emily~S. Cross}.} \bibinfo{year}{2024}\natexlab{b}.
\newblock \showarticletitle{Opening Up to Social Robots: How Emotions Drive Self-Disclosure Behavior}.
\newblock \bibinfo{journal}{\emph{arXiv preprint arXiv:2402.01023}} (\bibinfo{year}{2024}).
\newblock
\newblock
\shownote{TODO(camera-ready): verify final published venue}.


\bibitem[Leite et~al\mbox{.}(2013)]%
        {leite2013social}
\bibfield{author}{\bibinfo{person}{Iolanda Leite}, \bibinfo{person}{Carlos Martinho}, {and} \bibinfo{person}{Ana Paiva}.} \bibinfo{year}{2013}\natexlab{}.
\newblock \showarticletitle{Social Robots for Long-Term Interaction: A Survey}.
\newblock \bibinfo{journal}{\emph{International Journal of Social Robotics}} \bibinfo{volume}{5}, \bibinfo{number}{2} (\bibinfo{year}{2013}), \bibinfo{pages}{291--308}.
\newblock


\bibitem[Lewis et~al\mbox{.}(2020)]%
        {lewis2020retrieval}
\bibfield{author}{\bibinfo{person}{Patrick Lewis}, \bibinfo{person}{Ethan Perez}, \bibinfo{person}{Aleksandra Piktus}, \bibinfo{person}{Fabio Petroni}, \bibinfo{person}{Vladimir Karpukhin}, \bibinfo{person}{Naman Goyal}, \bibinfo{person}{Heinrich K{\"u}ttler}, \bibinfo{person}{Mike Lewis}, \bibinfo{person}{Wen-tau Yih}, \bibinfo{person}{Tim Rockt{\"a}schel}, \bibinfo{person}{Sebastian Riedel}, {and} \bibinfo{person}{Douwe Kiela}.} \bibinfo{year}{2020}\natexlab{}.
\newblock \showarticletitle{Retrieval-Augmented Generation for Knowledge-Intensive {NLP} Tasks}. In \bibinfo{booktitle}{\emph{Proceedings of NeurIPS}}.
\newblock


\bibitem[Lipner et~al\mbox{.}(2022)]%
        {lipner2022rupture}
\bibfield{author}{\bibinfo{person}{Lauren~M. Lipner}, \bibinfo{person}{J.~Christopher Muran}, \bibinfo{person}{Catherine~F. Eubanks}, \bibinfo{person}{Bernard~S. Gorman}, {and} \bibinfo{person}{Arnold Winston}.} \bibinfo{year}{2022}\natexlab{}.
\newblock \showarticletitle{Operationalizing Alliance Rupture-Repair Events Using Control Chart Methods}.
\newblock \bibinfo{journal}{\emph{Clinical Psychology \& Psychotherapy}} \bibinfo{volume}{29}, \bibinfo{number}{1} (\bibinfo{year}{2022}), \bibinfo{pages}{339--350}.
\newblock


\bibitem[Matsuyama et~al\mbox{.}(2016)]%
        {matsuyama2016socially}
\bibfield{author}{\bibinfo{person}{Yoichi Matsuyama}, \bibinfo{person}{Arjun Bhardwaj}, \bibinfo{person}{Ran Zhao}, \bibinfo{person}{Oscar Romeo}, \bibinfo{person}{Sushma Akoju}, {and} \bibinfo{person}{Justine Cassell}.} \bibinfo{year}{2016}\natexlab{}.
\newblock \showarticletitle{Socially-Aware Animated Intelligent Personal Assistant Agent}. In \bibinfo{booktitle}{\emph{Proceedings of SIGDIAL}}. \bibinfo{pages}{224--227}.
\newblock
\href{https://doi.org/10.18653/v1/W16-3628}{doi:\nolinkurl{10.18653/v1/W16-3628}}


\bibitem[Meinshausen and B{\"u}hlmann(2010)]%
        {meinshausen2010stability}
\bibfield{author}{\bibinfo{person}{Nicolai Meinshausen} {and} \bibinfo{person}{Peter B{\"u}hlmann}.} \bibinfo{year}{2010}\natexlab{}.
\newblock \showarticletitle{Stability Selection}.
\newblock \bibinfo{journal}{\emph{Journal of the Royal Statistical Society: Series B (Statistical Methodology)}} \bibinfo{volume}{72}, \bibinfo{number}{4} (\bibinfo{year}{2010}), \bibinfo{pages}{417--473}.
\newblock


\bibitem[M{\"u}ller et~al\mbox{.}(2018)]%
        {muller2018detecting}
\bibfield{author}{\bibinfo{person}{Philipp M{\"u}ller}, \bibinfo{person}{Michael~Xuelin Huang}, {and} \bibinfo{person}{Andreas Bulling}.} \bibinfo{year}{2018}\natexlab{}.
\newblock \showarticletitle{Detecting Low Rapport During Natural Interactions in Small Groups from Non-Verbal Behaviour}. In \bibinfo{booktitle}{\emph{Proceedings of the 23rd International Conference on Intelligent User Interfaces}}. \bibinfo{pages}{153--164}.
\newblock


\bibitem[Oertel et~al\mbox{.}(2020)]%
        {oertel2020engagement}
\bibfield{author}{\bibinfo{person}{Catharine Oertel}, \bibinfo{person}{Ginevra Castellano}, \bibinfo{person}{Mohamed Chetouani}, \bibinfo{person}{Jauwairia Nasir}, \bibinfo{person}{Mohammad Obaid}, \bibinfo{person}{Catherine Pelachaud}, {and} \bibinfo{person}{Christopher Peters}.} \bibinfo{year}{2020}\natexlab{}.
\newblock \showarticletitle{Engagement in Human-Agent Interaction: An Overview}.
\newblock \bibinfo{journal}{\emph{Frontiers in Robotics and AI}}  \bibinfo{volume}{7} (\bibinfo{year}{2020}).
\newblock


\bibitem[{OpenAI}(2025)]%
        {openai2025gpt41}
\bibfield{author}{\bibinfo{person}{{OpenAI}}.} \bibinfo{year}{2025}\natexlab{}.
\newblock \bibinfo{title}{Introducing {GPT-4.1} in the {API}}.
\newblock \bibinfo{howpublished}{\url{https://openai.com/index/gpt-4-1/}}.
\newblock
\newblock
\shownote{Accessed: 2026-03-22}.


\bibitem[Saeki et~al\mbox{.}(2024)]%
        {saeki-etal-2024-InteLLA}
\bibfield{author}{\bibinfo{person}{Mao Saeki}, \bibinfo{person}{Hiroaki Takatsu}, \bibinfo{person}{Fuma Kurata}, \bibinfo{person}{Shungo Suzuki}, \bibinfo{person}{Masaki Eguchi}, \bibinfo{person}{Ryuki Matsuura}, \bibinfo{person}{Kotaro Takizawa}, \bibinfo{person}{Sadahiro Yoshikawa}, {and} \bibinfo{person}{Yoichi Matsuyama}.} \bibinfo{year}{2024}\natexlab{}.
\newblock \showarticletitle{{I}nte{LLA}: Intelligent Language Learning Assistant for Assessing Language Proficiency through Interviews and Roleplays}. In \bibinfo{booktitle}{\emph{Proceedings of the 25th Annual Meeting of the Special Interest Group on Discourse and Dialogue}}. \bibinfo{publisher}{Association for Computational Linguistics}, \bibinfo{address}{Kyoto, Japan}, \bibinfo{pages}{385--399}.
\newblock
\href{https://doi.org/10.18653/v1/2024.sigdial-1.34}{doi:\nolinkurl{10.18653/v1/2024.sigdial-1.34}}


\bibitem[Saito and Rehmsmeier(2015)]%
        {saito2015precision}
\bibfield{author}{\bibinfo{person}{Takaya Saito} {and} \bibinfo{person}{Marc Rehmsmeier}.} \bibinfo{year}{2015}\natexlab{}.
\newblock \showarticletitle{The Precision-Recall Plot Is More Informative than the {ROC} Plot When Evaluating Binary Classifiers on Imbalanced Datasets}.
\newblock \bibinfo{journal}{\emph{PLoS ONE}} \bibinfo{volume}{10}, \bibinfo{number}{3} (\bibinfo{year}{2015}), \bibinfo{pages}{e0118432}.
\newblock


\bibitem[Santana et~al\mbox{.}(2025)]%
        {santana2025speechtojoy}
\bibfield{author}{\bibinfo{person}{Ricardo Santana}, \bibinfo{person}{Bahar Irfan}, \bibinfo{person}{Erik Lagerstedt}, \bibinfo{person}{Gabriel Skantze}, {and} \bibinfo{person}{Andre Pereira}.} \bibinfo{year}{2025}\natexlab{}.
\newblock \showarticletitle{Speech-to-Joy: Self-Supervised Features for Enjoyment Prediction in Human--Robot Conversation}. In \bibinfo{booktitle}{\emph{Proceedings of the 27th International Conference on Multimodal Interaction}}.
\newblock


\bibitem[Sidner et~al\mbox{.}(2005)]%
        {sidner2005explorations}
\bibfield{author}{\bibinfo{person}{Candace~L. Sidner}, \bibinfo{person}{Christopher Lee}, \bibinfo{person}{Cory~D. Kidd}, \bibinfo{person}{Neal Lesh}, {and} \bibinfo{person}{Charles Rich}.} \bibinfo{year}{2005}\natexlab{}.
\newblock \showarticletitle{Explorations in Engagement for Humans and Robots}.
\newblock \bibinfo{journal}{\emph{Artificial Intelligence}} \bibinfo{volume}{166}, \bibinfo{number}{1-2} (\bibinfo{year}{2005}), \bibinfo{pages}{140--164}.
\newblock


\bibitem[Skjuve et~al\mbox{.}(2022)]%
        {skjuve2022longitudinal}
\bibfield{author}{\bibinfo{person}{Marita Skjuve}, \bibinfo{person}{Asbj{\o}rn F{\o}lstad}, \bibinfo{person}{Knut~Inge Fostervold}, {and} \bibinfo{person}{Petter~Bae Brandtzaeg}.} \bibinfo{year}{2022}\natexlab{}.
\newblock \showarticletitle{A Longitudinal Study of Human--Chatbot Relationships}.
\newblock \bibinfo{journal}{\emph{International Journal of Human-Computer Studies}}  \bibinfo{volume}{168} (\bibinfo{year}{2022}), \bibinfo{pages}{102903}.
\newblock


\bibitem[Tickle-Degnen and Rosenthal(1990)]%
        {tickle1990nature}
\bibfield{author}{\bibinfo{person}{Linda Tickle-Degnen} {and} \bibinfo{person}{Robert Rosenthal}.} \bibinfo{year}{1990}\natexlab{}.
\newblock \showarticletitle{The Nature of Rapport and Its Nonverbal Correlates}.
\newblock \bibinfo{journal}{\emph{Psychological Inquiry}} \bibinfo{volume}{1}, \bibinfo{number}{4} (\bibinfo{year}{1990}), \bibinfo{pages}{285--293}.
\newblock


\bibitem[Tsakalidis et~al\mbox{.}(2021)]%
        {tsakalidis2021patterns}
\bibfield{author}{\bibinfo{person}{Adam Tsakalidis}, \bibinfo{person}{Dana Atzil-Slonim}, \bibinfo{person}{Asaf Polakovski}, \bibinfo{person}{Natalie Shapira}, \bibinfo{person}{Rivka Tuval-Mashiach}, {and} \bibinfo{person}{Maria Liakata}.} \bibinfo{year}{2021}\natexlab{}.
\newblock \showarticletitle{Automatic Identification of Ruptures in Transcribed Psychotherapy Sessions}. In \bibinfo{booktitle}{\emph{Proceedings of the Seventh Workshop on Computational Linguistics and Clinical Psychology (CLPsych)}}. ACL, \bibinfo{pages}{122--128}.
\newblock


\bibitem[Yamazaki et~al\mbox{.}(2023)]%
        {yamazaki2023}
\bibfield{author}{\bibinfo{person}{Yamazaki} {et~al\mbox{.}}} \bibinfo{year}{2023}\natexlab{}.
\newblock \showarticletitle{TODO: verify title (Yamazaki et al., 2023 in-home relational robot study, N=13)}.
\newblock
\newblock
\shownote{TODO(camera-ready): verify authors, title, and venue}.


\bibitem[Zhao et~al\mbox{.}(2016)]%
        {zhao2016rapport}
\bibfield{author}{\bibinfo{person}{Ran Zhao}, \bibinfo{person}{Tanmay Sinha}, \bibinfo{person}{Alan~W. Black}, {and} \bibinfo{person}{Justine Cassell}.} \bibinfo{year}{2016}\natexlab{}.
\newblock \showarticletitle{Socially-Aware Virtual Agents: Automatically Assessing Dyadic Rapport from Temporal Patterns of Behavior}. In \bibinfo{booktitle}{\emph{Proceedings of the 16th International Conference on Intelligent Virtual Agents}}. \bibinfo{pages}{218--233}.
\newblock


\end{thebibliography}
